\documentclass{article}

\usepackage[ruled]{algorithm2e}
\SetKwInOut{Input}{Input}
\SetKwInOut{Output}{Output} 
\newcommand\algcom[1]{\tcp{#1}}

\usepackage{amsfonts,amsmath}
\usepackage{caption}
\usepackage{subcaption}
\usepackage{url}

\usepackage{tikz}
\tikzset{
        MyPersp/.style={scale=1.8,
                        x={(-0.8cm,-0.4cm)},
                        y={(0.8cm,-0.4cm)},
                        z={(0cm,1cm)}},
        MyPersp2/.style={scale=2.2,
                         x={(-0.8cm,-0.4cm)},
                         y={(0.8cm,-0.4cm)},
                         z={(0cm,1cm)}},
        PerspYZ/.style={scale=1.8,
                        x={(0cm,0cm)},
                        y={(1cm,0cm)},
                        z={(0cm,1cm)}},
        MyPoints/.style={fill=white,draw=black,thick},
        BluePoints/.style={fill=white,draw=blue!50,thick}
}

\usetikzlibrary{calc,fadings,decorations.pathreplacing}

\usetikzlibrary{positioning}

\newcommand\pgfmathsinandcos[3]{%
  \pgfmathsetmacro#1{sin(#3)}%
  \pgfmathsetmacro#2{cos(#3)}%
}
\newcommand\LongitudePlane[3][current plane]{%
  \pgfmathsinandcos\sinEl\cosEl{#2} 
  \pgfmathsinandcos\sint\cost{#3} 
  \tikzset{#1/.estyle={cm={\cost,\sint*\sinEl,0,\cosEl,(0,0)}}}
}
\newcommand\LatitudePlane[3][current plane]{%
  \pgfmathsinandcos\sinEl\cosEl{#2} 
  \pgfmathsinandcos\sint\cost{#3} 
  \pgfmathsetmacro\yshift{\cosEl*\sint}
  \tikzset{#1/.estyle={cm={\cost,0,0,\cost*\sinEl,(0,\yshift)}}} %
}

\usetikzlibrary{shadows}

\usepackage{gensymb}

\usepackage{xspace}

\newcommand\ie{\emph{i.e.}\xspace}

\newcommand\etal{\emph{et~al.}\xspace}

\title{Real-time Correction of Panoramic Images using Hyperbolic M{\"o}bius
        Transformations}

\author{Luis~Pe{\~n}aranda
\thanks{IMPA -- Instituto de Matem{\'a}tica Pura e Aplicada, Rio de
Janeiro, Brazil}
\thanks{UFRJ -- Universidade Federal do Rio de Janeiro, Brazil}\\
luisp@impa.br
\and Luiz~Velho
\footnotemark[1]\\
lvelho@impa.br
\and Leonardo~Sacht
\footnotemark[1]
\thanks{UFSC -- Universidade Federal de Santa Catarina, Florian{\'o}polis,
Brazil}\\
leo-ks@impa.br}

\begin{document}
\maketitle

\begin{abstract}
        Wide-angle images gained a huge popularity in the last years due to
        the development of computational photography and imaging technological
        advances. They present the information of a scene in a way which is
        more natural for the human eye but, on the other hand, they
        introduce artifacts such as bent lines. These artifacts become more
        and more unnatural as the field of view increases.

        In this work, we present a technique aimed to improve the
        perceptual quality of panorama visualization. The main ingredients
        of our approach are, on one hand, considering the viewing sphere as
        a Riemann sphere, what makes natural the application of M{\"o}bius
        (complex) transformations to the input image, and, on the other
        hand, a projection scheme which changes in function of the field of
        view used.

        We also introduce an
        implementation of our method, compare it against images produced
        with other methods and show that the transformations can be done in
        real-time, which makes our technique very appealing for new
        settings, as well as for existing interactive panorama
        applications.


\end{abstract}

\section{Introduction}

An image is called panoramic when it represents a wide field of view (FOV,
in photography, is the part of the world which is visible through the camera).

Some applications generate panoramic images from an image representing the
\emph{viewing sphere}, a sphere containing the scene, centered at the
viewpoint. The viewing sphere is usually obtained with special cameras,
software or a combination of both. A panorama can be obtained by
\emph{stitching} together a set of regular pictures of a scene, producing
an image inscribed on the surface of a sphere, and then the sphere
projected on a plane. The entire process can be done by using programs like
Hugin~\cite{hugin}. This software uses state-of-the-art methods to find
common points in different images of the same scene, combine those images,
correct colors and project the sphere. In the last years, however, cameras
capable of directly obtaining a 360-degree sphere became popular; these
cameras directly output a spherical image, without the need of special
software to combine pictures~\cite{phgb11}. There also exists the
well-known fisheye lens~\cite{k89}, which permits to capture a 180-degree
image directly on a plane.

There exists a number of transformation techniques to obtain a panoramic
image from the viewing sphere (typically, a projection from the sphere to a
plane). Each type of transformation has distinct properties, and the goal
of using different transformations is usually to obtain a more realistic
image. The subjective concept of \emph{realistic} can be interpreted in
different ways, typically as bending straight lines as less as possible or
as preserving angles of the scene. Since lines and angles cannot be
preserved at the same time~\cite{zb95}, different projection schemes were
proposed. Warping techniques were also developed~\cite{caa09,k10}; however,
they need long human interaction or optimization methods and
thus cannot be used for real-time nor interactive applications.

While one kind of transformation might be good for some setting, it might
be bad for other. This situation becomes evident on some interfaces where a
user is able to interactively change the FOV of a scene. Usually,
perspective or equirectangular projections (both known for providing good
results for small to medium FOVs) are used for interactive visualizations.
When the FOV becomes very wide, projections are perceptually less realist.

What we propose in this paper is to adopt a new approach in interactive
applications for obtaining a panoramic image when the FOV becomes wide.
When the user widens the FOV until surpassing a limit where common
projections tend to produce bad results, we propose to simulate the
widening of the FOV by performing a M{\"o}bius transformation, and then
apply the initially intended common projection. The novelty of this
approach is that we do not introduce a complicated projection method but,
instead, we perform a transformation on the viewing sphere and then we
project using a well-known perspective projection. Performing
transformations of the image directly on the viewing sphere is specially
important in some settings, on which it is not desired to project on a
plane (for instance, when projecting on a dome) and, as far as we know, is
also a novelty of our method.

Our main contribution is a projection method which performs a
transformation on the viewing sphere, which makes it applicable on
non-plane projections. Moreover, our method can be used in real-time
applications. We also use a powerful tool such as complex transformations
for our purpose, showing that they can be employed in Image Processing and
open the quest for new applications in the field. Finally, we present an
open-source implementation of the method.

A preliminary version of this work was published in~\cite{sv13}. The
present full-version includes the theoretical foundations of the method,
new experiments (comprising comparison with more state-of-the-art methods),
argues about perceptual properties of the generated images, adds a
discussion on the applications of the new technique and motivates future
research. Additionally, the text was entirely rewritten.

The roadmap of the paper is as follows. Next Section argues about previous
work on the field. Section~\ref{sec:math} introduces some mathematical
definitions to understand the rest of the paper. Our approach is formalized
in Section~\ref{sec:our}.
Sections~\ref{subsec:perceptual}~and~\ref{sec:impl} provide an analysis of
the technique, while Section~\ref{sec:appl} introduces some scenarios where
our technique can be applied. Finally, Section~\ref{sec:future} discusses
current work and future research directions.

\section{Previous Work}

Due to their increasing popularity and interest, panoramic images have
become a theme of intense discussion in the Computer Graphics and Image
Processing communities in the last twenty years. The impossibility of
obtaining a global projection from the sphere to the plane that preserves
all possible straight lines and object shapes was shown in the seminal work
by Zorin and Barr~\cite{zb95}. This theoretical limitation motivated much
research for obtaining perceptually realistic panoramas.

One approach for this problem was to use different perspective projections
in the same scene~\cite{zm05}. The user specifies different projection
planes and view directions to define the different projections.  The
discontinuities caused by using different projections for different regions
of the panorama were hidden (if possible) by choosing the projection planes
in a way that fit well orientation discontinuities that were already
present in the scene.

Other approach consisted in investigating near-perspective projections used
by ancient painters, such as Pannini projection~\cite{spg10}. This
technique produces very good results, as proved in some ancient paintings.

Conformal mappings were also explored with the aim of preserving the shape
of the objects in a panoramic image~\cite{gbdpps07}. This approach
consisted in investigating the stereographic projection and scaling of the
complex plane, but only for artistic and exploratory purposes. Since the
focus was on shape preservation, results present bent lines. In our work,
we go beyond and improve these ideas to map wide fields of view to narrower
ones and add a perspective re-projection step, which produces better
quality results because the straight lines are less bent.

Other methods relied on both user interaction and energy-minimization
formulations. Carroll~\etal~\cite{caa09} used the important lines in the
scene provided by the user and detected faces to control straight line
preservation and conformality in these regions. Kopf~\etal~\cite{kldcc09}
used regions specified by the user where the projection should be nearly
planar to formulate their optimization framework. 
Wei~\etal~\cite{wlhmt12} formulates the problem as the minimization of a
quadratic energy based on  user annotation, which has a closed form
and implies in the solution of a sparse linear system.
In these approaches, user
interaction was usually laborious and the optimization formulations made
them impossible to be implemented in real-time\footnote{For us,
\emph{real-time} means that the computations are done quicker than the
rendering of a frame in the application. Even if the other methods are
fast, they are not suitable to be implemented in real-time.}.

An interesting approach consisted in dynamically changing the projection
used depending on the FOV~\cite{kudc07}. This can also be achieved by our
viewer by applying different shrink values
as the FOV of the perspective projection changes, as explained in
Section~\ref{sec:our}. Also, our
visualization simulates better camera movements since it is a natural
generalization of the perspective projection.

Another advantage of our method compared to previous ones is that we do not
rely on heavy user interaction. The user is only asked to vary
one parameter (which may be set automatically to a good value),
what makes our panorama viewer a pleasant experience, instead of laborious.
Also, our formulation is simple and does not rely on heavy optimizations,
which makes possible to implement our method in real-time.

\section{Definitions}
\label{sec:math}

\begin{figure}
\centering
\begin{tikzpicture} 


\def\R{2.5} 
\def\angEl{29} 
\def\angAz{-105} 
\def\angPhi{-40} 
\def\angBeta{19} 


\pgfmathsetmacro\H{\R*cos(\angEl)} 
\tikzset{xyplane/.estyle={cm={cos(\angAz),sin(\angAz)*sin(\angEl),-sin(\angAz),
                              cos(\angAz)*sin(\angEl),(0,0)}}}
\LongitudePlane[xzplane]{\angEl}{\angAz}
\LongitudePlane[pzplane]{\angEl}{\angPhi}
\LatitudePlane[equator]{\angEl}{0}


        \foreach \t in {0,15,...,165}
        {\draw[MyPersp2,gray] ({cos(\t)},{sin(\t)},0)
                \foreach \rho in {5,10,...,360}
                {--({cos(\t)*cos(\rho)},{sin(\t)*cos(\rho)},
                                {sin(\rho)})}--cycle;
        }
        \foreach \t in {-75,-60,...,75}
        {\draw[MyPersp2,gray] ({cos(\t)},0,{sin(\t)})
                \foreach \rho in {5,10,...,360}
                {--({cos(\t)*cos(\rho)},{cos(\t)*sin(\rho)},
                                {sin(\t)})}--cycle;
        }




\coordinate (O) at (0,0);
\coordinate (N) at (0,\H);
\coordinate (S) at (0,-\H);
\path[pzplane] (\angEl:\R) coordinate (Phat);
\path[pzplane] (2*\R,0) coordinate (PE);
\path[xzplane] (\R,0) coordinate (XE);
\path (PE) ++(0,0) coordinate (Paux); 
\coordinate (P) at (intersection cs: first line={(N)--(Phat)},
                                        second line={(S)--(Paux)});
\path[equator] (-43.3:\R) coordinate (P0);






\draw[dashed] (Phat) -- (N);
\fill[MyPoints] (O) circle (1pt) node[below left,black] {$O$};
\draw (Phat) -- (P);
\fill[MyPoints] (P) circle (1pt) node[above right] {$p$};
\fill[MyPoints] (Phat) circle (1pt) node[above right] {$\hat{p}$};
\draw[dashed] (O) -- (P0);
\draw (P0) -- (P);
\fill[MyPoints] (N) circle (1pt) node[above left] {$N$};



\end{tikzpicture}
\caption{Correspondence between a point \(\hat{p}\) on the Riemann sphere
and a point \(p\) on the complex plane.}
\label{fig:riemann}
\end{figure}
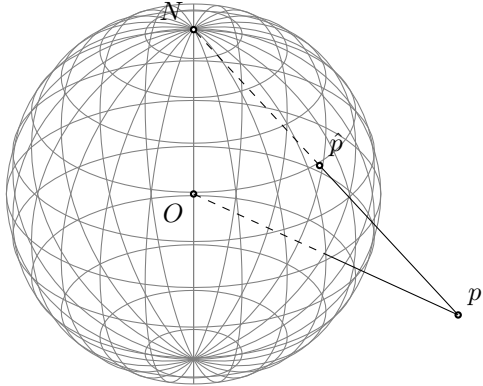

Before describing our method, we formalize in this Section the ideas
presented so far. We assume the given image is inscribed in the viewing
sphere. The viewing sphere will be considered a Riemann sphere. Each point
on the sphere
corresponds thus to the representation of a complex number;
Figure~\ref{fig:riemann} shows the correspondence between points in the
complex plane and on the Riemann sphere. A
complex number \(p=x+iy\) is a point on the complex plane, with coordinates
\(x\) and \(y\). This plane is the stereographic projection of the Riemann
sphere. The point \(p\) in the complex plane is represented on the sphere
by the point \(\hat{p}\). \(p\) and \(\hat{p}\) are related by
stereographic equations.
A complete study of stereographic projections is beyond the scope of the
present work. We will only show the formulas of stereographic projections
here and refer to Snyder's book~\cite{s87} for details.

The stereographic projection maps complex points in points on the viewing
sphere. It is defined as follows.
\begin{equation}
\begin{array}{rrcl}
       \mathbf{S}:
       &\mathbb{S}^2\setminus \{(0,0,-1)\}
       &\rightarrow
       &\mathbb{C}\\
       &\hat{p}(\hat{x},\hat{y},\hat{z})
       &\mapsto
       &p\left( \frac{2\hat{x}}{\hat{z}+1}, \frac{2\hat{y}}{\hat{z}+1} \right)
\end{array}
\end{equation}

On the other hand, the inverse stereographic projections maps points
on the viewing sphere to complex points, and is defined as follows.
\begin{equation}
\begin{array}{rrcl}
       \mathbf{S}^{-1}:
       &\mathbb{C}
       &\rightarrow
       &\mathbb{S}^2\setminus \{(0,0,-1)\}\\
       &p(x,y)
       &\mapsto
       &\hat{p}\left(   \frac{4x}{x^2+y^2+4},
                        \frac{4y}{x^2+y^2+4},
                        \frac{x^2+y^2-4}{x^2+y^2+4} \right)
\end{array}
\end{equation}

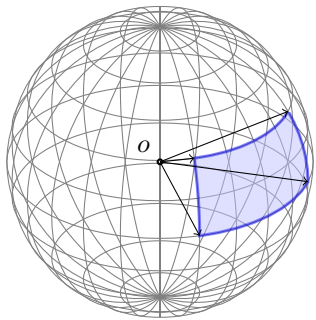
\begin{figure}
\centering
\begin{tikzpicture}[MyPersp]

        \foreach \t in {0,15,...,165}
        {\draw[gray] ({cos(\t)},{sin(\t)},0)
                \foreach \rho in {5,10,...,360}
                {--({cos(\t)*cos(\rho)},{sin(\t)*cos(\rho)},
                                {sin(\rho)})}--cycle;
        }
        \foreach \t in {-75,-60,...,75}
        {\draw[gray] ({cos(\t)},0,{sin(\t)})
                \foreach \rho in {5,10,...,360}
                {--({cos(\t)*cos(\rho)},{cos(\t)*sin(\rho)},
                                {sin(\t)})}--cycle;
        }

        \draw[blue,very thick,opacity=0.5,fill=blue!25]
                ({cos(60)},{sin(60)},0)
                \foreach \rho in {5,10,...,30}
                {--({cos(60)*cos(\rho)},{sin(60)*cos(\rho)},{sin(\rho)})}
                \foreach \rho in {67.5,75,...,120}
                {--({cos(30)*cos(\rho)},{cos(30)*sin(\rho)},{sin(30)})}
                \foreach \rho in {30,25,...,5}
                {--({cos(120)*cos(\rho)},{sin(120)*cos(\rho)},{sin(\rho)})}
                \foreach \rho in {120,112.5,...,67.5}
                {--({cos(0)*cos(\rho)},{cos(0)*sin(\rho)},{sin(0)})}
                --cycle;

        \coordinate (orig) at (0,0,0);
        \fill[MyPoints] (orig) circle (.5pt)node[above left]{$o$};
        \draw[->] (orig) -- ({cos(0)*cos(60)},{cos(0)*sin(60)},{sin(0)});
        \draw[->] (orig) -- ({cos(0)*cos(120)},{cos(0)*sin(120)},{sin(0)});
        \draw[->] (orig) -- ({cos(30)*cos(60)},{cos(30)*sin(60)},{sin(30)});
        \draw[->] (orig) -- ({cos(30)*cos(120)},{cos(30)*sin(120)},{sin(30)});

\end{tikzpicture}
\caption{The viewing sphere. The FOV (blue region) is to be mapped to a
rectangular image. \(o\) is the observers position, in the center of the
sphere.}
\label{fig:sphere}
\end{figure}

We assume the observer stands in the center of the viewing sphere. She
can do two basic things: rotate her head in any direction or change the
FOV. For the sake of simplicity they will be studied separately, but it would
not require much extra work to consider them together.

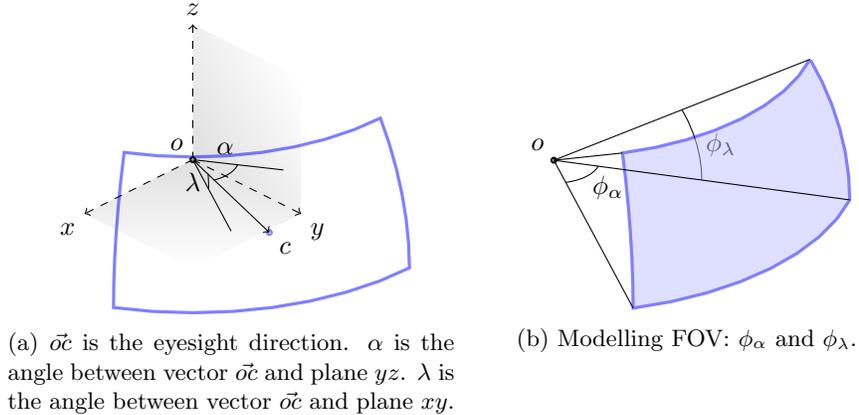
\begin{figure}
\centering
\begin{subfigure}{.49\textwidth}
\centering
\begin{tikzpicture}[MyPersp,font=\normalsize,scale=2]

        \coordinate (orig) at (0,0,0);
        \coordinate (xa) at (.5,0,0);
        \coordinate (ya) at (0,.5,0);
        \coordinate (za) at (0,0,.5);

        \shade[top color=gray!25,bottom color=transparent!0,shading angle=135]
                (orig)--(0,0,.5)--(0,.5,.5)--(ya)--cycle;
        \shade[top color=gray!25,bottom color=transparent!0]
                (orig)--(ya)--(.5,.5,0)--(.5,0,0)--cycle;

        \draw (orig)--({.5*cos(15)*cos(60)},{.5*cos(15)*sin(60)},0);
        \draw (orig)--(0,{.5*cos(15)*sin(60)},{.5*sin(15)});

        \draw[dashed,->] (orig) -- (xa);
        \node[below left] at (xa) {$x$};
        \draw[dashed,->] (orig) -- (ya);
        \node[below right] at (ya) {$y$};
        \draw[dashed,->] (orig) -- (za);
        \node[above] at (za) {$z$};

        \draw ({.2*cos(60)*cos(0)},{.2*sin(60)*cos(0)},{.2*sin(0)})
                \foreach \rho in {5,7,...,15}
                {--({.2*cos(60)*cos(\rho)},
                    {.2*sin(60)*cos(\rho)},
                    {.2*sin(\rho)})};
        \node at (.1,.1,0) {$\lambda$};

        \draw ({.15*cos(0)*cos(122)},{.15*cos(0)*sin(122)},{.15*sin(0)})
                \foreach \rho in {120,118,...,72}
                {--({.15*cos(0)*cos(\rho)},
                    {.15*cos(0)*sin(\rho)},
                    {.15*sin(0)})};
        \node at (0,.15,.1) {$\alpha$};

        \draw[blue,very thick,opacity=0.5]
                ({cos(30)},{sin(30)},0)
                \foreach \rho in {5,10,...,30}
                {--({cos(30)*cos(\rho)},{sin(30)*cos(\rho)},{sin(\rho)})}
                \foreach \rho in {37.5,45,...,90}
                {--({cos(30)*cos(\rho)},{cos(30)*sin(\rho)},{sin(30)})}
                \foreach \rho in {30,25,...,5}
                {--({cos(90)*cos(\rho)},{sin(90)*cos(\rho)},{sin(\rho)})}
                \foreach \rho in {90,82.5,...,37.5}
                {--({cos(0)*cos(\rho)},{cos(0)*sin(\rho)},{sin(0)})}
                --cycle;

        \fill[MyPoints] (orig) circle (.25pt)node[above left]{$o$};
        \coordinate (cblue) at ({cos(15)*cos(60)},{cos(15)*sin(60)},{sin(15)});
        \fill[BluePoints] (cblue) circle (.25pt)node[below right]{$c$};
        \draw[->] (orig) -- (cblue);

\end{tikzpicture}
\caption{\(\vec{oc}\) is the eyesight direction. \(\alpha\) is the angle
between vector \(\vec{oc}\) and plane \(yz\). \(\lambda\) is the angle
between vector \(\vec{oc}\) and plane \(xy\).}
\label{subfig:rot}
\end{subfigure}
\begin{subfigure}{.49\textwidth}
\centering
\begin{tikzpicture}[MyPersp,font=\normalsize,scale=2]

        \draw ({.15*cos(0)*cos(120)},{.15*cos(0)*sin(120)},{.15*sin(0)})
                \foreach \rho in {120,112.5,...,60}
                {--({.15*cos(0)*cos(\rho)},{.15*cos(0)*sin(\rho)},{.15*sin(0)})};

        \node at
                ({.25*cos(0)*cos(90)},{.25*cos(0)*sin(90)},{.25*sin(0)})
                {$\phi_\alpha$};

        \draw ({.5*cos(120)*cos(30)},{.5*sin(120)*cos(30)},{.5*sin(30)})
                \foreach \rho in {25,20,...,0}
                {--({.5*cos(120)*cos(\rho)},{.5*sin(120)*cos(\rho)},{.5*sin(\rho)})};

        \node[right] at
                ({.5*cos(120)*cos(15)},{.5*sin(120)*cos(15)},{.5*sin(15)})
                {$\phi_\lambda$};

        \draw[blue,very thick,opacity=0.5,fill=blue!25]
                ({cos(60)},{sin(60)},0)
                \foreach \rho in {5,10,...,30}
                {--({cos(60)*cos(\rho)},{sin(60)*cos(\rho)},{sin(\rho)})}
                \foreach \rho in {67.5,75,...,120}
                {--({cos(30)*cos(\rho)},{cos(30)*sin(\rho)},{sin(30)})}
                \foreach \rho in {30,25,...,5}
                {--({cos(120)*cos(\rho)},{sin(120)*cos(\rho)},{sin(\rho)})}
                \foreach \rho in {120,112.5,...,67.5}
                {--({cos(0)*cos(\rho)},{cos(0)*sin(\rho)},{sin(0)})}
                --cycle;

        \coordinate (orig) at (0,0,0);
        \fill[MyPoints] (orig) circle (.25pt)node[above left]{$o$};
        \draw (orig) -- ({cos(0)*cos(60)},{cos(0)*sin(60)},{sin(0)});
        \draw (orig) -- ({cos(0)*cos(120)},{cos(0)*sin(120)},{sin(0)});
        \draw (orig) -- ({cos(30)*cos(60)},{cos(30)*sin(60)},{sin(30)});
        \draw (orig) -- ({cos(30)*cos(120)},{cos(30)*sin(120)},{sin(30)});

\end{tikzpicture}
\caption{Modelling FOV: \(\phi_\alpha\) and \(\phi_\lambda\).}
\label{subfig:fov}
\end{subfigure}
\caption{Details of the viewing sphere and the view region.}
\label{fig:spheredetail}
\end{figure}

The rotation of the observer's head is represented as two angles,
\(\alpha\) (azimuth) and \(\lambda\) (altitude). These angles represent
uniquely a point on the sphere, which is associated with the center of the
produced image, see Figure~\ref{subfig:rot}.

The FOV is again represented as two angles, \(\phi_\alpha\) (azimuthal FOV)
and \(\phi_\lambda\) (altitudinal FOV), as depicted in
Figure~\ref{subfig:fov}. However, the ratio between these two angles is
defined by the aspect ratio of the output image.

The problem consists then in obtaining a plane image of the FOV (the part
shaded in blue in Figure~\ref{fig:sphere}) in function of \(\alpha\),
\(\lambda\), \(\phi_\alpha\) and \(\phi_\lambda\).

The parameters \(\alpha\), \(\lambda\) and \(\phi_\alpha\)
(\(\phi_\lambda\) can be considered a function of \(\phi_\alpha\))
determine the points on the sphere to be processed. Recall that each point
on the sphere is uniquely determined by \(\phi_\lambda\) and
\(\phi_\alpha\) (because the radius of the sphere is fixed) or by its
cartesian coordinates in the original plane.

To study the variation of \(\phi_\alpha\), we assume \(\alpha\) and
\(\lambda\) fixed to zero. This way, a value of \(\phi_\alpha\) uniquely
determines the four points on the sphere which are the vertices of the blue
region of Figure~\ref{fig:sphere}.
For a given \(\phi_\alpha\), the FOV is computed, and points on the sphere
are mapped to the output image by a projection function (which usually
depends on the FOV).

Changes in viewing directions are given by changing \(\alpha\) and
\(\lambda\). Since the FOV does not change, these changes can be modeled as
rotations of the sphere. This approach has the advantage that the
projection function does not need to be recomputed, although
it must be composed with a rotation function on the sphere.

Finally, let us mention that the perspective projection is a key ingredient
of our method. Since it is a well-known projection method, we will not
describe it, but only mention that it is defined by the following equation.

\begin{equation}
\label{eq:persp}
\begin{array}{rrcl}
       \mathbf{P}:
       &\mathbb{S}^2 \setminus \{ z<0 \}
       &\rightarrow
       &\mathbb{C}\\
       &\hat p(x,y,z)
       &\mapsto
       &p\left( \frac x z , \frac y z \right)
\end{array}
\end{equation}

In practice, Equation~\ref{eq:persp} is scaled, in order for a given FOV to
be projected in the extents of an image. In the sequel, we will denote a
function performing a perspective projection of a FOV \(\phi\) as
\(\mathbf{P}_\phi\). We refer the reader to the book by Snyder~\cite{s87}
for more details and a complete study of perspective projections.

\section{The M{\"o}bius Projection Method}
\label{sec:our}

This Section introduces our method, aimed to visualize images with wide
FOVs. Perspective projections work very well in practice for small to
medium FOVs. When \(\phi_\alpha\) grows to a big value, it becomes
difficult to conceive a good projection function. One approach is to
replace the perspective projection \(\mathbf{P}_{\phi_\alpha}\) by a
warping which respects visual restrictions, such as conformality and
preservation of straight lines, as much as possible. We refer to~\cite{k10}
for details on such methods. The drawback of these methods is that they are
slow and they usually need human interaction, what makes them unsuitable
for interactive applications.

Our approach consists in using two different projection methods, depending
on the desired FOV. To the best of our knowledge, the only work on which
the projection scheme used varies as a function of the FOV is the paper by
Kopf~\etal~\cite{kudc07}. They use an adaptive projection resulting from
an interpolation between a perspective and a cylindrical projection.

We assume that there exists a value \(\phi_{\text{max}}\), such that the
perspective projection does not introduce big distortions when
\(\phi_\alpha \le \phi_{\text{max}}\). When this happens, we propose to use
a direct perspective projection, that is, without performing any shrink.

\begin{figure}
\centering
\begin{tikzpicture}[MyPersp,font=\normalsize]

        \foreach \t in {0,15,...,165}
        {\draw[gray] ({cos(\t)},{sin(\t)},0)
                \foreach \rho in {5,10,...,360}
                {--({cos(\t)*cos(\rho)},{sin(\t)*cos(\rho)},
                                {sin(\rho)})}--cycle;
        }
        \foreach \t in {-75,-60,...,75}
        {\draw[gray] ({cos(\t)},0,{sin(\t)})
                \foreach \rho in {5,10,...,360}
                {--({cos(\t)*cos(\rho)},{cos(\t)*sin(\rho)},
                                {sin(\t)})}--cycle;
        }

        \draw[blue,very thick,opacity=0.5,fill=blue!25]
                ({cos(15)*cos(-7.5)},{sin(15)*cos(-7.5)},{sin(-7.5)})
                \foreach \rho in {-2.5,2.5,...,37.5}
                {--({cos(15)*cos(\rho)},{sin(15)*cos(\rho)},{sin(\rho)})}
                \foreach \rho in {22.5,30,...,105}
                {--({cos(37.5)*cos(\rho)},{cos(37.5)*sin(\rho)},{sin(37.5)})}
                \foreach \rho in {32.5,27.5,...,-7.5}
                {--({cos(105)*cos(\rho)},{sin(105)*cos(\rho)},{sin(\rho)})}
                \foreach \rho in {105,98.5,...,15}
                {--({cos(-7.5)*cos(\rho)},{cos(-7.5)*sin(\rho)},{sin(-7.5)})}
                --cycle;

        \draw[red,very thick,opacity=0.5,fill=red!25]
                ({cos(30)},{sin(30)},0)
                \foreach \rho in {5,10,...,30}
                {--({cos(30)*cos(\rho)},{sin(30)*cos(\rho)},{sin(\rho)})}
                \foreach \rho in {37.5,45,...,90}
                {--({cos(30)*cos(\rho)},{cos(30)*sin(\rho)},{sin(30)})}
                \foreach \rho in {30,25,...,5}
                {--({cos(90)*cos(\rho)},{sin(90)*cos(\rho)},{sin(\rho)})}
                \foreach \rho in {90,82.5,...,37.5}
                {--({cos(0)*cos(\rho)},{cos(0)*sin(\rho)},{sin(0)})}
                --cycle;

        \draw[->] ({cos(15)*cos(-7.5)},{sin(15)*cos(-7.5)},{sin(-7.5)})
                -- ({cos(30)},{sin(30)},0);
        \draw[->] ({cos(15)*cos(37.5)},{sin(15)*cos(37.5)},{sin(37.5)})
                -- ({cos(30)*cos(30)},{sin(30)*cos(30)},{sin(30)});
        \draw[->] ({cos(37.5)*cos(105)},{cos(37.5)*sin(105)},{sin(37.5)})
                --({cos(30)*cos(90)},{cos(30)*sin(90)},{sin(30)});
        \draw[->] ({cos(105)*cos(-7.5)},{sin(105)*cos(-7.5)},{sin(-7.5)})
                --({cos(90)*cos(0)},{sin(90)*cos(0)},{sin(0)});
\end{tikzpicture}
\caption{The M{\"o}bius transformation, shrinking points in the new FOV (in
blue) to the old FOV (in red).}
\label{fig:mobius}
\end{figure}
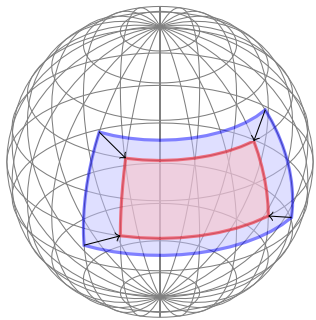

On the other hand, when \(\phi_\alpha > \phi_{\text{max}}\), the
perspective projection fails to provide a realistic image. We propose, in
this case, to perform a hyperbolic M{\"o}bius transformation~\cite{n97} on
the sphere points, in order to map points of the new FOV region to the old
FOV region, as depicted in Figure~\ref{fig:mobius} (this is achieved
by first mapping the sphere points to the complex plane, then applying 
the M{\"o}bius transformation to the complex plane and, finally, lifting 
the points back to the sphere). After, the same projection
\(\mathbf{P}_{\phi_{\text{max}}}\) maps points on the sphere to the output
image.

Our projection schema is non-conformal and does not preserve areas (it
inherits these properties from the perspective projection). In addition,
since hyperbolic M{\"o}bius transformations do not preserve great circles on the Riemann sphere, our
method looses the line-preservation property from perspective projections.
We will show in Section~\ref{sec:impl} that,
despite these properties, our method produces indeed very realistic images.

\subsection{Computing the Shrink}
\label{subsec:shrink}

As mentioned earlier, the problem is to find the M{\"o}bius transformation
that shrinks the new FOV into the old FOV, as depicted in
Figure~\ref{fig:mobius}. If we had the point \(c\) of
Figure~\ref{subfig:rot} coinciding with one of the poles, we could simply
apply an hyperbolic transformation to shrink the FOV. This can be
accomplished by rotating the sphere and making \(c\) coincide with the
south pole (corresponding to the origin on the complex plane).
Mathematically, this means to do a translation by a vector
\(\mathbf{t}_\text{rot}(-\alpha,-\lambda)\) in polar coordinates (note that
this translation must not be done in the equirectangular image). Without
loss of generality, we assume in the rest of the paper that
\(\alpha=\lambda=0\).

The following step is to perform a shrink centered in the origin. This can
be accomplished by using a hyperbolic M{\"o}bius transformation. The only
knowledge about M{\"o}bius transformations needed to understand the method
is that the shrink transformation we need takes the form \(M_{s}=\rho z\).
We refer the reader to Appendix~\ref{app:mobius} for some more details on
M{\"o}bius transformations.

At this point, after being introduced to M{\"o}bius transformations, the
reader should ask why we emphasize that our method uses them, while it is
only a polar scaling. The reason is that we believe this kind of
transformations has a myriad of applications in Image Processing, and we
want to encourage further research on them.

It remains to compute \(\rho\). Since \(\phi_{\text{max}}\) and \(\phi\)
are angles, it is direct to compute, in the equirectangular domain, \(\rho
= \frac{\phi_{\text{max}}}{\phi}\). But \(M_s\) is defined for the points
on the complex plane: this means that a stereographic projection \(S\) must
be first applied to the points on the sphere, then the old FOV is mapped to
the new FOV using \(M_s\) and, finally, an inverse stereographic projection
\(S^{-1}\) is used to map back points to the sphere. This means that the
shrink transformation takes the form \(\mathbf{t}_\text{shrink} = S^{-1}
\circ M_s \circ S\). (Note that, if \(\alpha\) and \(\lambda\) are not
assumed to be zero, the translation \(\mathbf{t}_\text{rot}\) must be
applied before \(\mathbf{t}_\text{shrink}\), and the the inverse
translation \(-\mathbf{t}_\text{rot}\) must be applied after.)

\begin{figure}
\centering
\resizebox{\linewidth}{!}{
\begin{tikzpicture}[
        imgstyle/.style={
                rectangle,
                minimum height=15mm,
                minimum width=20mm,
                thin,draw=black,
                top color=white,bottom color=blue!30}]

        \node (eq) [imgstyle]
                {\begin{tabular}{c}
                        equirectangular\\image\\(sphere)
                \end{tabular}};
        \node (rot) [imgstyle,right=of eq]
                {\begin{tabular}{c}rotated\\sphere\end{tabular}};
        \node (sh) [imgstyle,right=of rot]
                {\begin{tabular}{c}contracted\\rotated\\sphere\end{tabular}};
        \node (rotcontr) [imgstyle,right=of sh]
                {\begin{tabular}{c}contracted\\sphere\end{tabular}};
        \node (proj) [imgstyle,right=of rotcontr]
                {\begin{tabular}{c}resulting\\projection\end{tabular}};
        \draw[->] (eq) -- (rot) node[midway,above]
                {\(\mathbf{t}_\text{rot}\)};
        \draw[->] (rot) -- (sh) node[midway,above]
                {\(\mathbf{t}_\text{shrink}\)};
        \draw[->] (sh) -- (rotcontr) node[midway,above]
                {\(-\mathbf{t}_\text{rot}\)};
        \draw[->] (rotcontr) -- (proj) node[midway,above]
                {\(\mathbf{P}_{\phi_{\text{max}}}\)};
\end{tikzpicture}
}
\caption{Pipeline of the shrinking process. Recall that
\(\mathbf{t}_\text{shrink} = S^{-1} \circ M_s \circ S\).}
\label{fig:pipeline}
\end{figure}
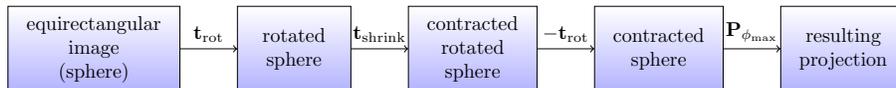

Finally, the desired projection is obtained by applying the transformation
\(\mathbf{P}_{\phi_{\text{max}}}\) to the resulting shrink. The pipeline of
the process is shown in Figure~\ref{fig:pipeline}.

The shrink is a good approach to reach wide FOVs on panoramas.
However, even with the shrink/projection scheme, a FOV close to \(2\pi\)
will produce an unrealistic output image, as well as introducing
unacceptable distortions. It becomes interesting thus to determine the
biggest \(\phi\) that can be visualized with this technique, as well as
computing a good value for \(\phi_{\text{max}}\). The first of these shall
be determined by experimentation, while the second can be approximated by a
function of \(\phi\), as well as determined experimentally. The next
Section addresses the computation of \(\phi_{\text{max}}\).

\section
{A Perceptually-Good Value of \(\phi_{\text{max}}\)}
\label{subsec:perceptual}

In this Section, we will argue about the value of the parameter
\(\phi_\text{max}\). As mentioned earlier, a value of \(\phi_{\text{max}}\)
can be determined experimentally, given a value of \(\phi\).

We will show first that our M{\"o}bius shrinking scheme
produces the same distortion than the stereographic projection by setting
\(\phi_\text{max}=\epsilon\), a small value.
As explained in Section~\ref{subsec:shrink}, given the image inscribed in
the viewing sphere, our method \(\mathbf{M}\) can be seen as a composition
of functions (we will assume again, without loss of generality, that the
sphere does not need to be rotated, in order to simplify the explanation).

\begin{equation}
\label{eq:MeqS}
\begin{array}{rccr}
        \mathbf{M}
        =& \mathbf{P}_\epsilon          & \circ &
                \multicolumn{1}{c}{\mathbf{t}_\text{shrink}} \\
        =& \mathbf{P}_\epsilon          & \circ & S^{-1} \circ M_s \circ S \\
        \approx& S                      & \circ & S^{-1} \circ M_s \circ S \\
        =&                              &       & M_s \circ S \\
        =&                              &       & S
\end{array}
\end{equation}

The approximation (in the third line of the equation) is due to the fact
that \(\epsilon\) is very small and, in this case, the perspective
projection \(\mathbf{P}_\epsilon\) is almost equal to the stereographic
projection \(S\). The last equality is forced by the fact that \(M_s \circ
S\) means to apply the stereographic projection to the sphere and then
shrink the resulting image; since this shrink does not deform the image, we
conclude that, in this setting, \(M_s\) has no effect.
This result will be corroborated experimentally in Section~\ref{sec:impl}.

Preservation of lengths and angles is important for many applications such
as cartography, but the stereographic projection is not ideal for panorama
visualization since it does not take into account perceptual aspects. In
the sequel we propose a choice of \(\phi_\text{max}\) that approximates
better human perception of the world.

It is known that the way images are formed on the human retina is very
similar to a perspective projection. Furthermore, our eyes only see with
clarity objects that are in the region called \emph{central}~FOV, which
comprises 60~degrees or less~\cite{cvshe13,w04}. We use this fact to
propose the value
\(
        \phi_{\text{max}} = 60 \degree \text{.}
\)
Using such value will map points on the portion of the viewing sphere
to be visualized to a 60 degrees FOV and apply a perspective projection
to this limited FOV, simulating then what one would see if was immersed in
the M{\"o}bius transformed sphere. 

\subsection{Discussion}

We stress that the choice \(\phi_\text{max} = 60\degree\) does not
guarantee straight lines in the scene to be straight in the final result.
This is due to the fact that the M{\"o}bius transformation on the sphere
does not map segments of great circles (projection of straight lines on the
sphere) to segments of great circles. If the user wants straight lines to
be better preserved, it is still possible to set a higher value to
\(\phi_\text{max}\). On the other hand, diminishing this value results in a
better preservation of lengths and angles. This aspect is further discussed
in the Section~\ref{sec:impl} and in the accompanying video. 

Since lines are not preserved, it is of interest to question if there are
some lines which are preserved, and the perceptual implications of the
answer. To begin this study, let us clarify how pieces of straight lines
(segments) on the scene are represented on the viewing sphere. One segment
on the scene and the center of the sphere form a triangle. The intersection
of two sides of the triangle with the sphere determines two points. The
representation of the segment on the viewing sphere is the geodesic joining
these two intersection points. The triangle is contained on a plane, and
the intersection of this plane with the viewing sphere is a great circle.
In general, straight lines on the scene are represented on the viewing
sphere as great circles or arcs of great circles.

Under a hyperbolic M{\"o}bius transformation, some circles on the sphere
are preserved, and some not. The great circles passing by the north
pole\footnote{Recall that, in our method, the viewing sphere is rotated to
        make coincide the north pole with the center of stretching, and
        rotated back after the stretching. For simplicity, we will assume
        here that the shrinking occurs with the north pole as center,
        without considering rotations.}
remain unchanged under hyperbolic M{\"o}bius transformations. The circles
parallel to the Equator are mapped to different circles parallel to the
Equator. The rest of the circles lose their shape when applying hyperbolic
M{\"o}bius transformations. This analysis permits to see that, in our
method, straight lines are deformed at different extents depending on the
distance of their projections on the viewing sphere to the shrinking point.

It remains to discuss about how bending lines affect perception. It is
widely accepted that a projection is better when it bends lines less, but
let us postulate that this is not quite true in human perception. It is
known that the human vision is very complex, far beyond the scope of this
paper, and that the brain reconstructs the image captured by the eyes to
form a three-dimensional model of the reality. Historically, camera lenses
were manufactured to try to produce perceptually-good images. One
meaningful example on this are the fisheye lenses for very wide FOVs, in
which lines are bent. We believe in this respect that the line bending of
our method is perceptually-acceptable for very wide FOVs. This fact will
also be corroborated in the next Section.

Let us mention, to conclude the Section, that one approach to partially
reduce line bending is to consider cylindrical projections. Cylindrical
projections do not bend vertical lines, but the problem remains intact for
horizontal lines. A direction to explore in the future would be to combine
M{\"o}bius transformations with cylindrical projections, as
Kopf~\etal~\cite{kudc07} do with perspective and cylindrical projections.

\subsection{A Final Note on Distortion}

It would be interesting to also perform the analysis from an orthogonal
point of view, that is, a completely automated method which does not
require human interaction. As far as we know, the only method aimed to
quantify the quality of a plane projection of a sphere is due to
Milnor~\cite{m69}. Briefly, this methodology computes the \emph{scale} of
each pair of points on a projection (the scale is defined as the ratio
between the geodesic distance on the sphere and the projected distance of
them). The distortion is the logarithm of the ratio between the smallest
and the biggest \emph{scales} of the projection.

We used Sage~Math~\cite{sage} to implement the formulas of the Milnor
distortion. For our experiments, we used a set of points on a grid of
$500 \times 500$ points on the surface of the unit sphere. We repeated, for
FOVs between $1\degree$ and $180\degree$, the following experiment. We
considered all the pairs of distinct points of this set (inside our FOV).
We computed the projections (using many different projection methods,
including ours) of each point and, for each pair of points, we computed the
\emph{scale} under each projection method. This procedure let us compute
the Milnor distortion, for a given FOV, for all different types of
projections. We compared then the distortions for each FOV.
The results were not the expected: the quantitative Milnor distortion of
the images does not correspond to their perceptual quality. This fact opens
a research direction for the future: to conceive quantitative methods for
projections, consistent with the human perception.

One alternative to our experiment is to choose the set of points to compute
Milnor distortion on by studying the content of images. However, the
results will be associated with a specific image and will not be as general
as we wanted when designed our experiment. We believe it deserves a full
research paper to study quantitative distortions of specific images.

\section{Implementation and Experiments}
\label{sec:impl}

\begin{table}
\caption{Frame rates obtained by our technique while the user interacts,
for different mesh resolutions (varying on the rows) and different input
equi-rectangular image resolutions (varying on the columns).  These results
were generated in a screen resolution of 1024 $\times$ 768 pixels in a PC
with an Intel Xeon Quad Core 2.13GHz with 12 GB of RAM and a GeForce GTX
470 GPU.}
\label{table:performance} \centering
\renewcommand{\arraystretch}{1.3}
\normalsize
\begin{tabular*}{.81\textwidth}{@{\extracolsep{\fill}}ccc}
  \hline
  & 4000 $\times$ 2000 pixels & 8000 $\times$ 4000 pixels\\
  \hline
  200 $\times$ 200 vertices & 93 fps & 89 fps\\
  400 $\times$ 400 vertices & 85 fps & 84 fps\\
  800 $\times$ 800 vertices & 35 fps &  33 fps\\
    \hline
\end{tabular*}
\end{table}

\begin{algorithm}[t]
  \BlankLine
  \Input{coordinates $(x,y,z)$ of a point in a plane image, angles
  $\lambda$ and $\phi$ and a scale $s$ to scale the complex plane}
  \Output{coordinates $(x_t,y_t,z_t)$ of the point, transformed using the
  M{\"o}bius projection method}

  \BlankLine
  \algcom{rotation of $-\lambda$ on the $xz$ plane}
  $(x,y,z) \leftarrow \big(\cos(-\lambda) x - \sin(-\lambda) z,
                           y,
                           \sin(-\lambda) x + \cos(-\lambda) z\big)$;\\

  \BlankLine
  \algcom{rotation of $-\phi$ on the $xz$ plane}
  $(x,y,z) \leftarrow \big(x,
                           \cos(-\phi) y - \sin(-\phi) z,
                           \sin(-\phi) y + \cos(-\phi) z\big)$;\\

  \BlankLine
  \algcom{stereographic projection}
  $(u,v) \leftarrow \big(2 x / (1-z),2 y / (1-z)\big)$;\\

  \BlankLine
  \algcom{convert from Cartesian to polar coordinates}
  $(\rho,\theta) \leftarrow \big(\sqrt{u^2+v^2},\arctan(u,v)\big)$;\\

  \BlankLine
  \algcom{scaling the complex plane according to $s$}
  $\rho \leftarrow \rho s$;\\

  \BlankLine
  \algcom{mapping back from polar to Cartesian coordinates}
  $(u,v) \leftarrow \big(-\rho \sin(\theta),\rho \cos(\theta)\big)$;\\

  \BlankLine
  \algcom{mapping back from the complex plane to the unit sphere}
  $(x,y,z) \leftarrow \big(4 u/(u^2+v^2+4),
                           4 v/(u^2+v^2+4),
                           (u^2+v^2-4)/(u^2+v^2+4)\big)$;\\

  \BlankLine
  \algcom{perspective projection}
  $(x_t,y_t,z_t) \leftarrow \big(x/(-z),y/(-z),z\big)$;\\

  \BlankLine
  \Return $(x_t,y_t,z_t)$;
  \caption{\label{alg:mobius} M{\"o}bius projection method}
\end{algorithm}

While in theory our method shows advantages over direct projections, we
implemented it to see how it performs in practice and the perceptual
quality of the produced images. We coded in C++, using OpenGL to deal with
image operations and texture the surface of the viewing sphere, and Qt for
the interface, what asserts portability. All geometric transformations on
the vertices of the sphere can be performed in parallel, what made us
implement them as a GLSL vertex shader.

The input of the shader algorithm is the viewing sphere, with the
equirectangular image as a texture. Each one of the vertices forming the
triangular grid approximating the sphere is processed by the vertex shader,
computing its position on the projection plane (applying the operations
described in Figure~\ref{fig:pipeline} on Page~\pageref{fig:pipeline}).
Finally, the fragment shader then interpolates the texture values inside
the resulting triangles, forming the output image.
Algorithm~\ref{alg:mobius} shows the pseudocode of the algorithm we
implemented in the vertex shader. It should be noted that the shader works
with homogeneous points but we used, in the pseudocode, three-dimensional
points, since the homogeneous component always equals one in our case. In
addition, we can safely assume the $z$-coordinate of the input points to be
zero.

The implementation consists in a window
showing the image processed through our technique. The interface can be
appreciated in the video accompanying this paper. It permits
to pan the image with the mouse, as well as adjusting the FOV and the parameter
\(\phi_{\text{max}}\) with the mouse wheel (the latter holding the
Shift
key).
All these operations are done in real-time on a desktop PC, as illustrated
in Table~\ref{table:performance} on Page~\pageref{table:performance}. The
code is open-source and available at
\url{http://git.impa.br/luisp/panoramic/}.

\begin{figure}
\centering
\begin{subfigure}{.49\textwidth}
\centering
\includegraphics[width=\textwidth]{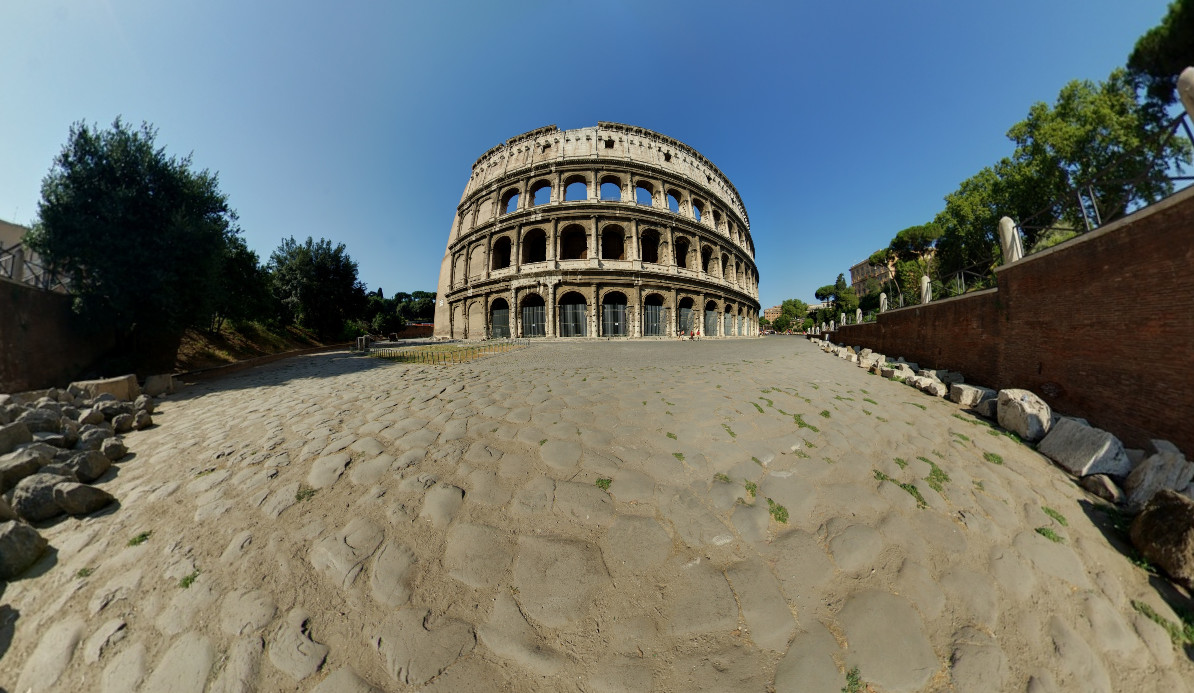}
\caption{Stereographic projection}
\label{subfig:fov240stereo}
\end{subfigure}
\begin{subfigure}{.49\textwidth}
\centering
\includegraphics[width=\textwidth]{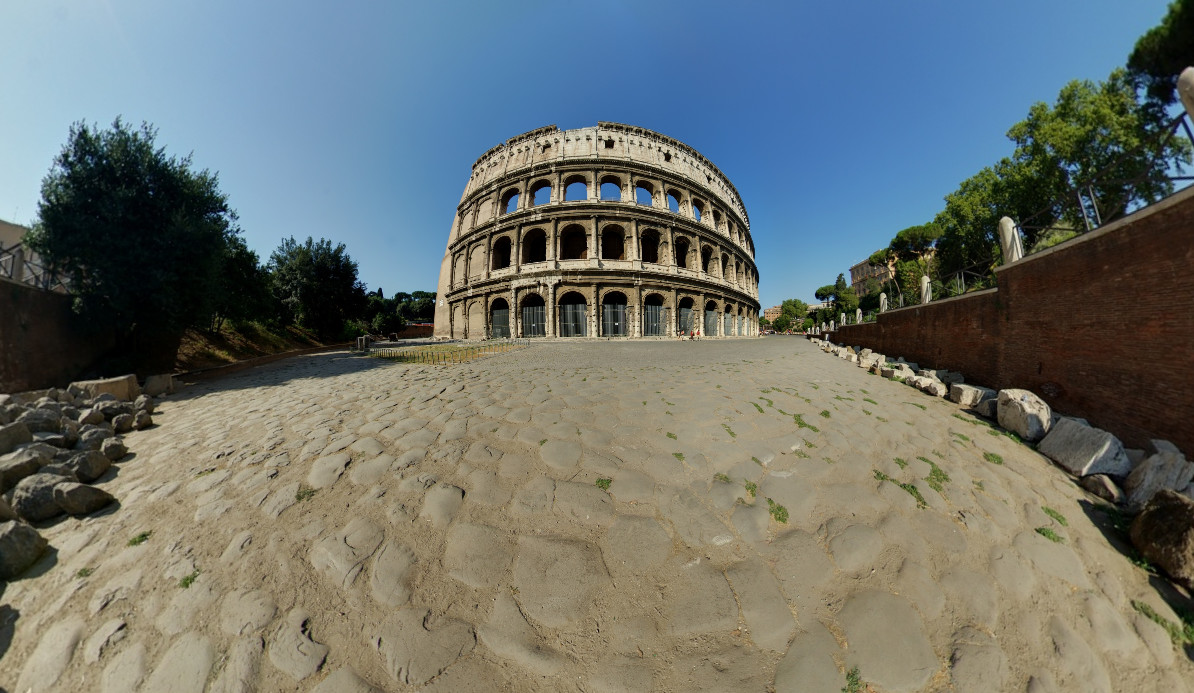}
\caption{M{\"o}bius, \(\phi_\text{max}=1\degree\approx 0.02\text{rad}\)}
\label{subfig:fov240fovmax1}
\end{subfigure}
\caption{The stereographic projection and the M{\"o}bius method with a
small value of \(\phi_\text{max}\) produce the same results, as predicted
in Equation~\ref{eq:MeqS}, on Section~\ref{subsec:perceptual}. In this
case, \(\phi=240\degree\approx 4.2\text{rad}\).}
\label{fig:fov240}
\end{figure}

We reproduce here some results of our tests. First of all, let us
corroborate the affirmation of Section~\ref{subsec:perceptual}, which
stated that our method produces similar images than the stereographic
projection. In Figure~\ref{fig:fov240}, the Roman Coliseum serves us to
show that the stereographic projection and our method produced images which
are indistinguishable for the human eye. This figure was produced using a
very wide FOV.

\begin{figure}
\centering
\begin{subfigure}{.49\textwidth}
\centering
\includegraphics[width=\textwidth]{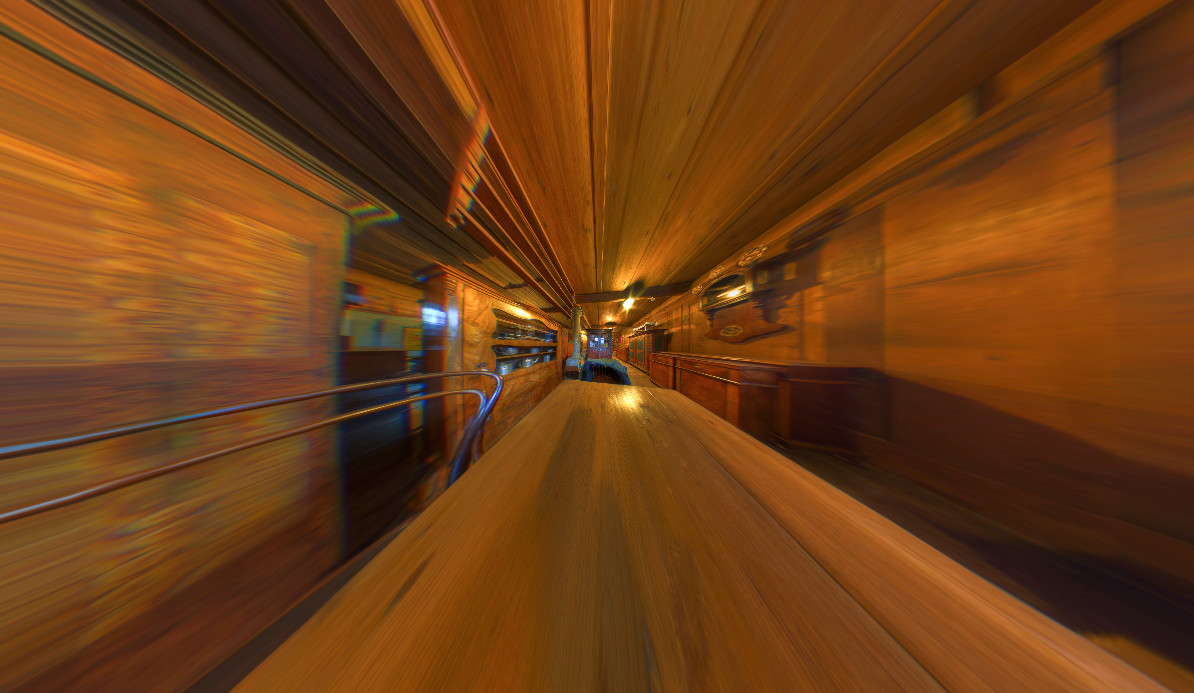}
\caption{Perspective}
\label{subfig:fov172persp}
\end{subfigure}
\begin{subfigure}{.49\textwidth}
\centering
\includegraphics[width=\textwidth]{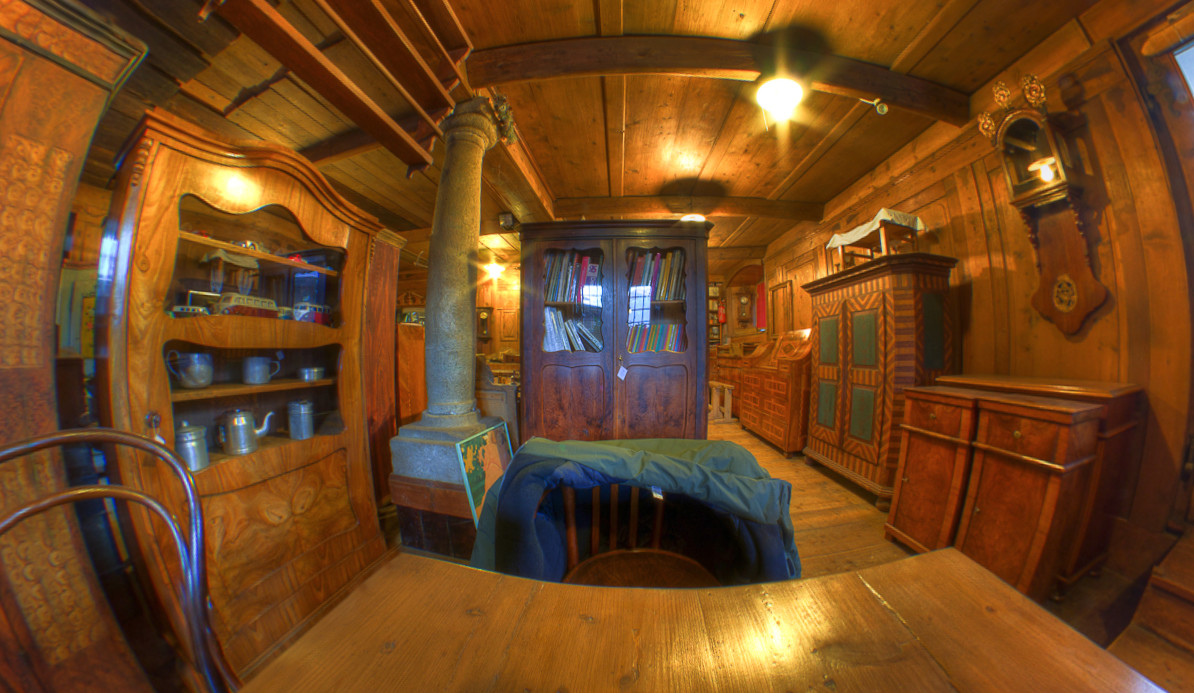}
\caption{Stereographic \(\approx\) M{\"o}bius, \(\phi_\text{max}=1\degree\)}
\label{subfig:fov172sterep}
\end{subfigure}
\begin{subfigure}{.49\textwidth}
\centering
\includegraphics[width=\textwidth]{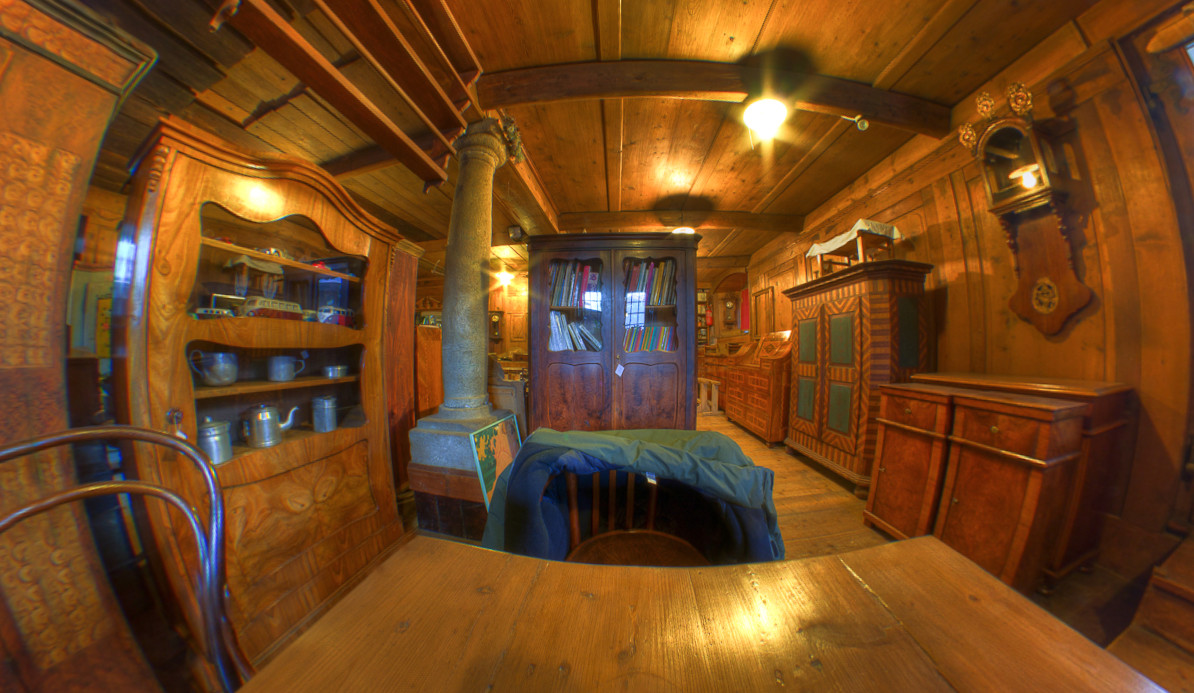}
\caption{M{\"o}bius, \(\phi_\text{max}=60\degree\)}
\label{subfig:fov172fovmax60}
\end{subfigure}
\begin{subfigure}{.49\textwidth}
\centering
\includegraphics[width=\textwidth]{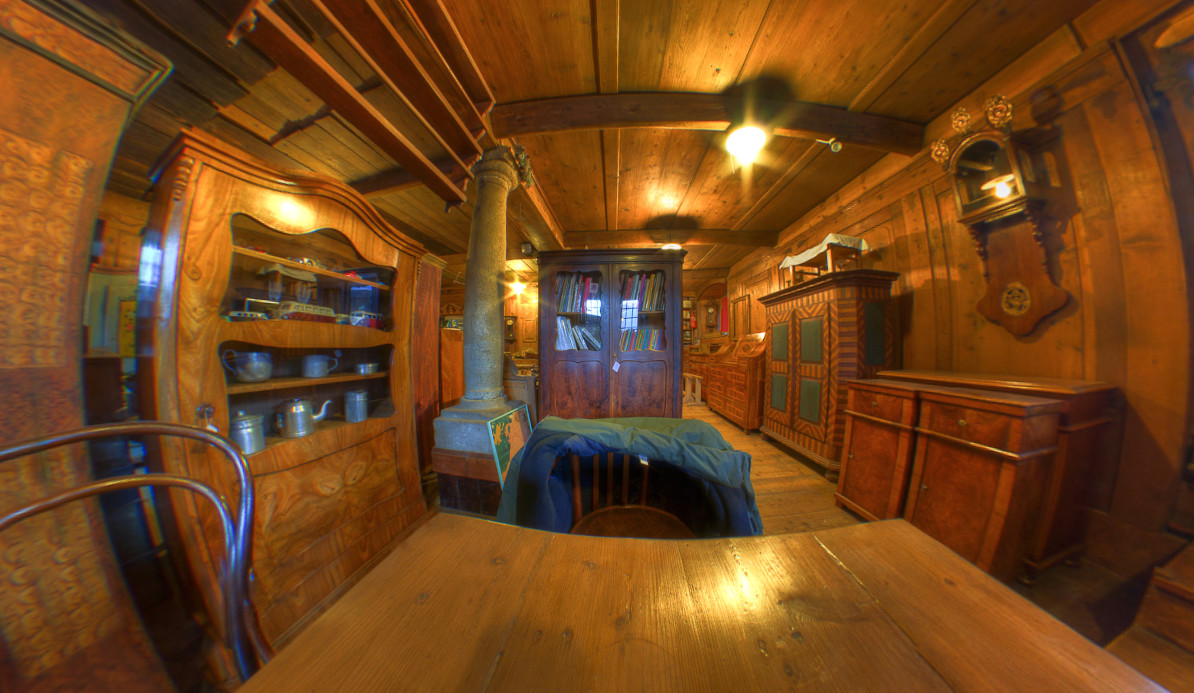}
\caption{M{\"o}bius, \(\phi_\text{max}=90\degree\)}
\label{subfig:fov172fovmax90}
\end{subfigure}
\caption{It is shown how the parameter \(\phi_\text{max}\) controls line
bending. In all subfigures, \(\phi=172\degree\approx 3\text{rad}\).}
\label{fig:fov172}
\end{figure}

Figure~\ref{fig:fov172} shows how our method improves the quality of the
images near the borders with respect to the perspective projection, while
keeping the image on the center undistorted. From this figure, as well as
from the video accompanying the paper, one can realize a meaning of the
parameter \(\phi_\text{max}\). In the figure, line bending is easily
distinguishable, since the scene is full of straight lines. Moreover, it is
evident how lines passing by the center of shrinking are preserved.

A big value of \(\phi_\text{max}\) (bigger than or equal to the FOV)
produces a perspective projection. Small values bend the lines on the
scene, but correct artifacts of the perspective projection. Smaller values
bend lines more and correct more artifacts. From the perceptual point of
view, bent lines are by far more acceptable than distorted images, as can
be appreciated in Figure~\ref{fig:fov172}. We can see \(\phi_\text{max}\)
as a parameter which controls the ratio between line straightness and
distortion near the borders. The price paid for obtaining images which are
clear close to the borders is, in this case, loosing the preservation of
straight lines.

\begin{figure}
  \centering
  \begin{tabular}{cc}
    \includegraphics[width=.47\textwidth]{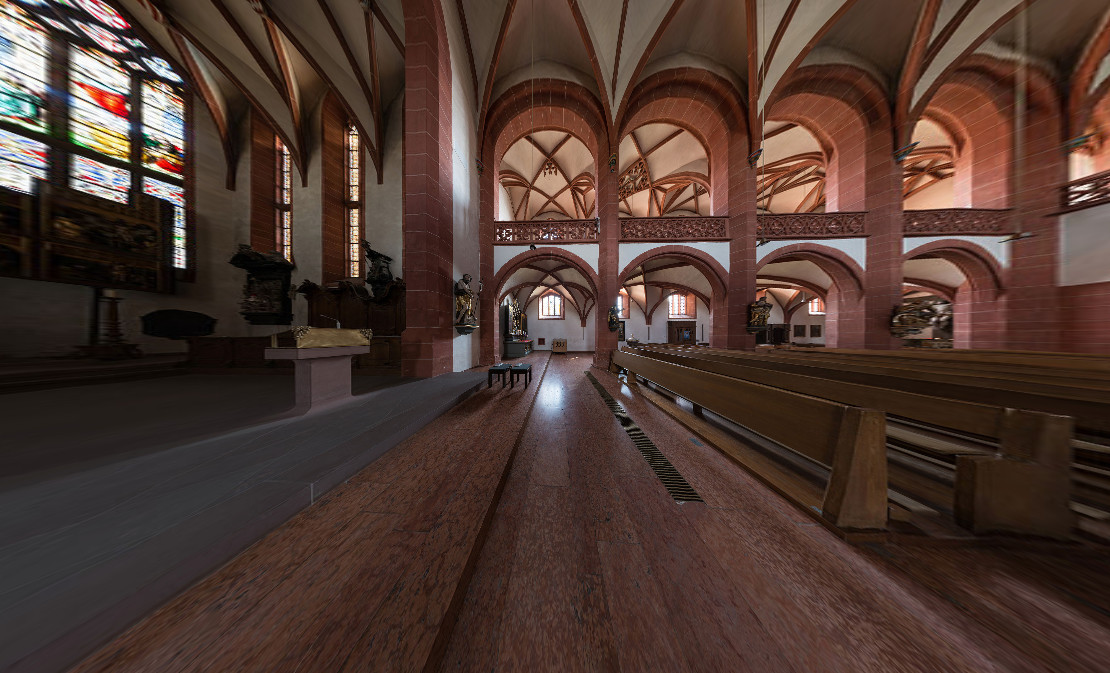} &
    \includegraphics[width=.47\textwidth]{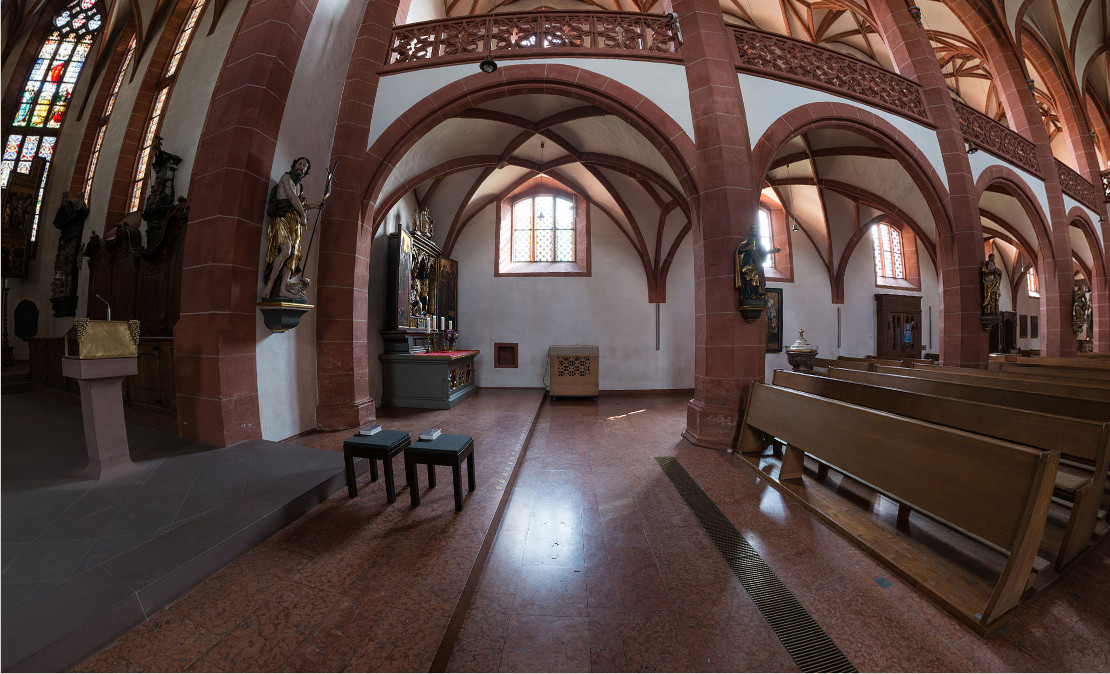}
    \\
    {\small (a) Perspective} &
    {\small (b) Stereographic}
    \\
  \end{tabular}
  \begin{tabular}{cc}

    \includegraphics[width=.47\textwidth]{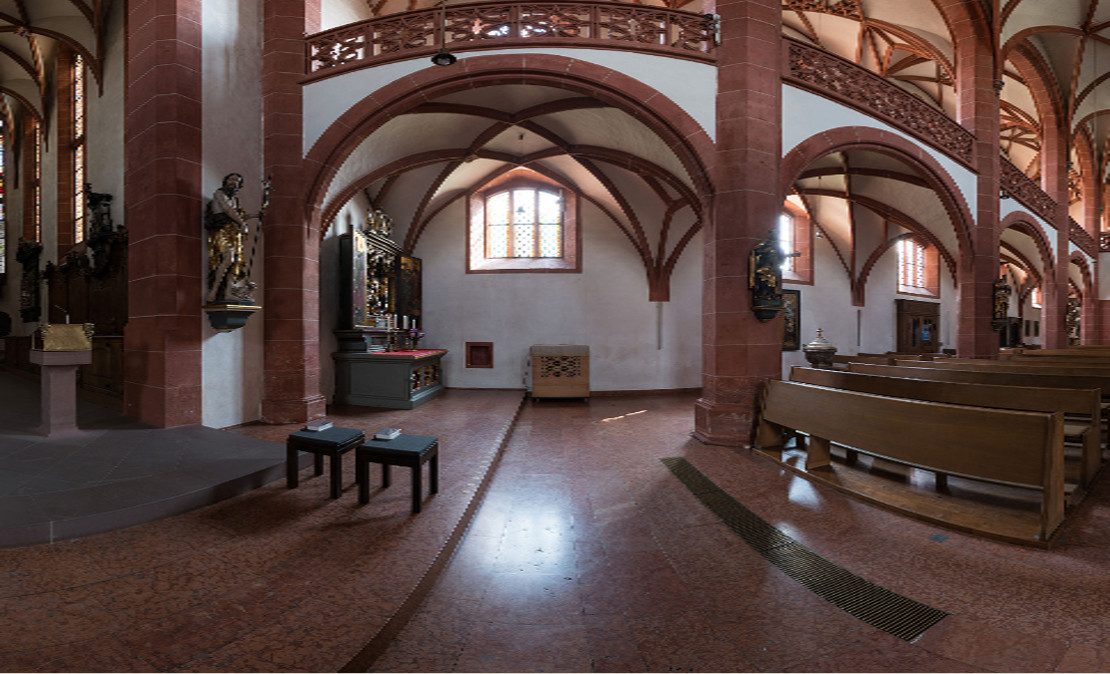} &
    \includegraphics[width=.47\textwidth]{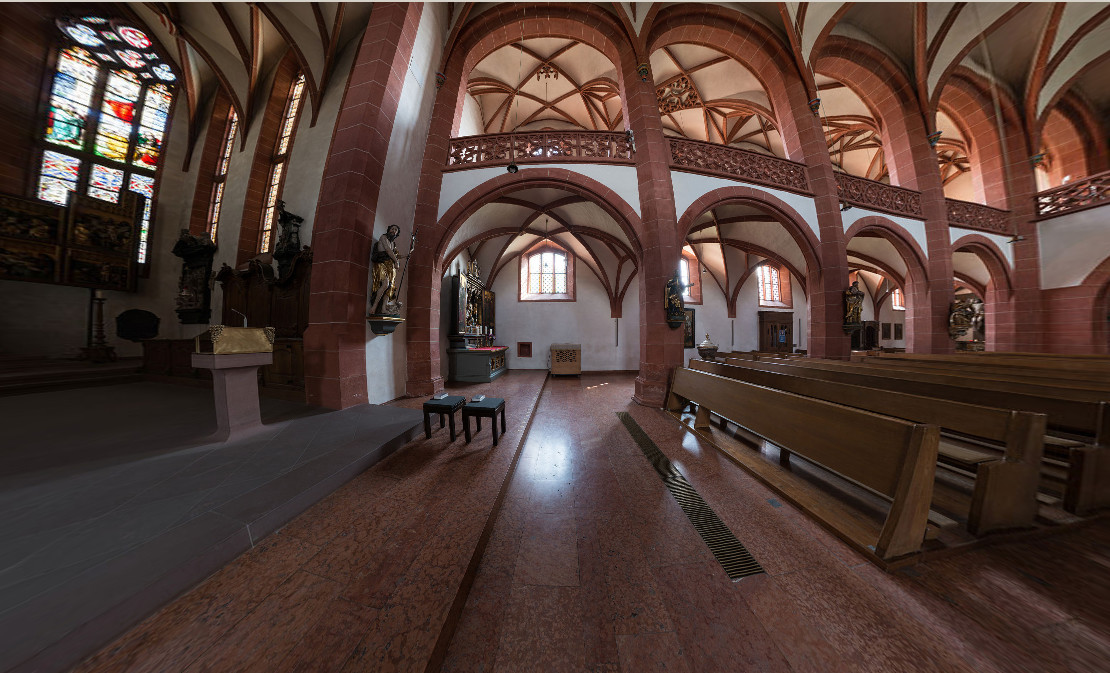}
    \\
    {\small (c) Mercator (cropped at \(157\degree\))} &
    {\small (d) Zorin-Barr, \(\lambda=0.2\)}
    \\
  \end{tabular}
  \begin{tabular}{cc}
    \includegraphics[width=.47\textwidth]{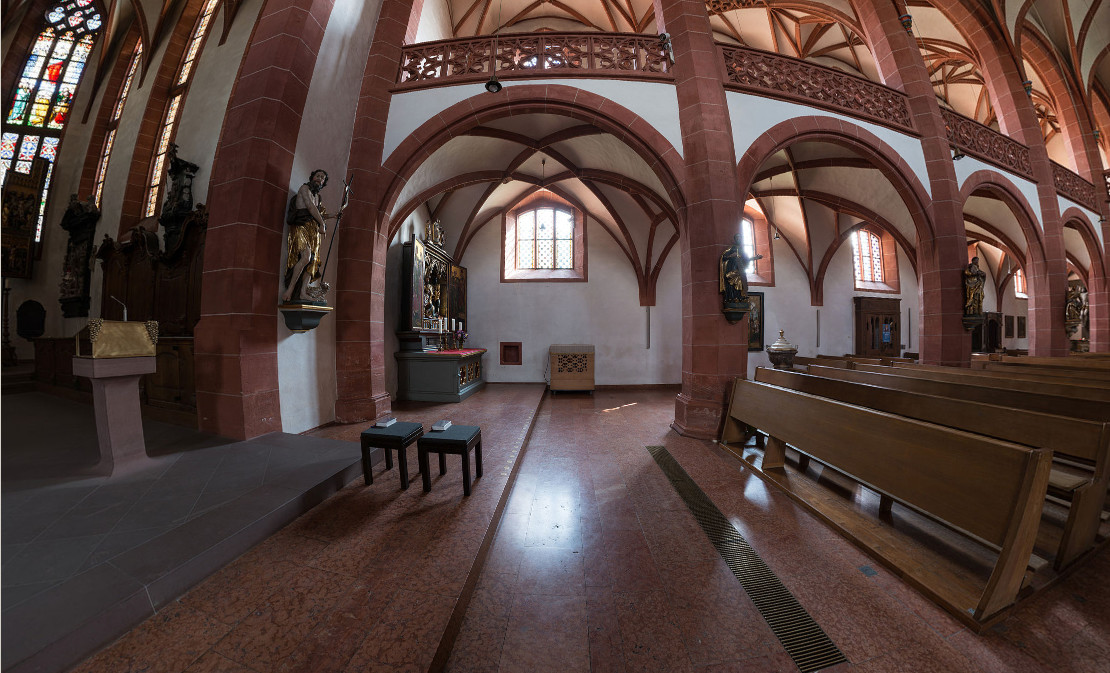} &
    \includegraphics[width=.47\textwidth]{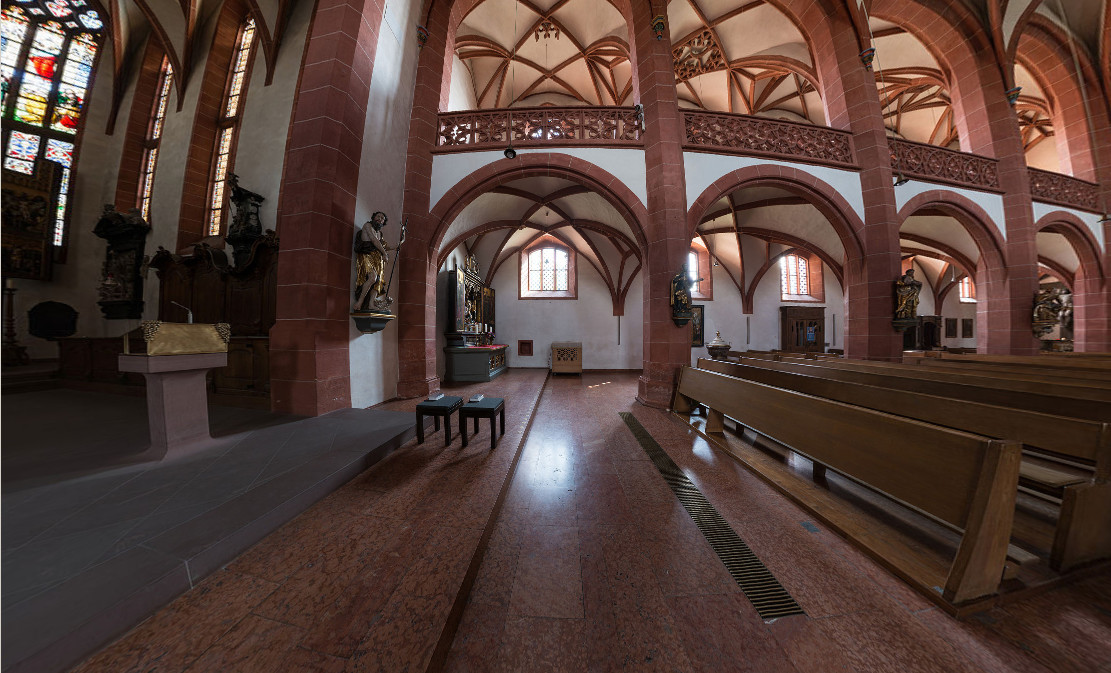}
    \\
    {\small (e) M{\"o}bius, \(\phi_\text{max}=60\degree\)} &
    {\small (f) M{\"o}bius, \(\phi_\text{max}=120\degree\)}
    \\
  \end{tabular}
  \caption{Comparison with standard projections for \(\phi = 160\degree\).
  Our method, (b), (e) and (f), has a good balance between straight line
  and object shape preservation, obtained by adjusting the parameter
  \(\phi_\text{max}\).}
  \label{fig:comparison_standard}
\end{figure}

We also compare our technique to previous methods in the field of
generation of panoramic images. We use first a view of the Rheingauer~Dom
to show in Figure~\ref{fig:comparison_standard} a comparison with standard
projections from the viewing sphere to the image plane. We observe that for
the field of view of 160 degrees the perspective projection
(Figure~\ref{fig:comparison_standard}a) distorts objects too much but
preserves all straight lines. The stereographic projection, equal to the
M{\"o}bius method when \(\phi_\text{max}=1\degree\),
(Figure~\ref{fig:comparison_standard}b) improves the perceptual quality of
the image by bending some lines. The Mercator projection
(Figure~\ref{fig:comparison_standard}c) preserves object shapes because it
is conformal, but bends much the lines. The perspective projection
corrected with the Zorin-Barr transformation, using parameters
\(\lambda=0.2\) and \(R=0.5\) (Figure~\ref{fig:comparison_standard}d),
yields an image with a reasonable straight line preservation, but presents
artifacts near the borders. The M{\"o}bius projection using the value of
\(\phi_\text{max}=60\degree\) (Figure~\ref{fig:comparison_standard}f)
produced a perceptually good result. Finally, we set the parameter
\(\phi_{\text{max}}\) to \(120\degree\)
(Figure~\ref{fig:comparison_standard}f) to obtain an image with less line
bending.

\begin{figure}
  \begin{tabular*}{\textwidth}{@{\extracolsep{\fill}}cc}
    \includegraphics[height=4.5cm]{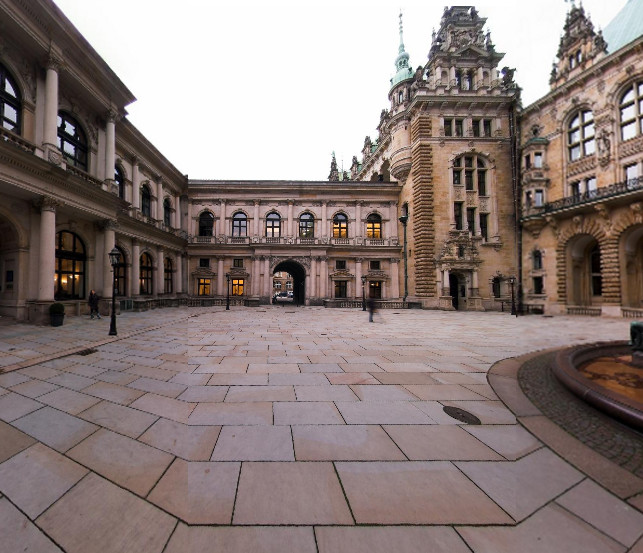}
    &
    \includegraphics[height=4.5cm]{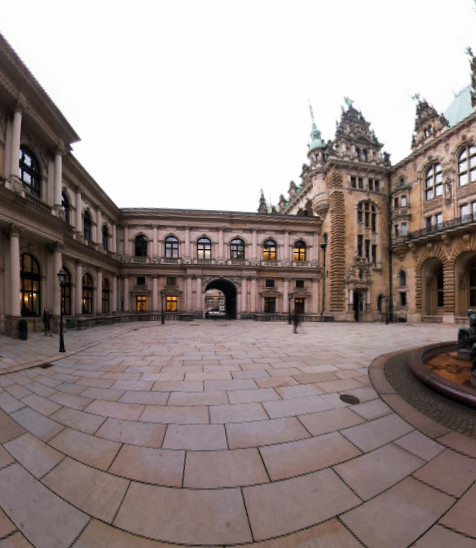}
    \\
    {\small (a) Zelnik-Manor~\etal~\cite{zm05}}
    &
    {\small (b) Carroll~\etal~\cite{caa09}}
    \\
    \\
    \includegraphics[height=4.5cm]{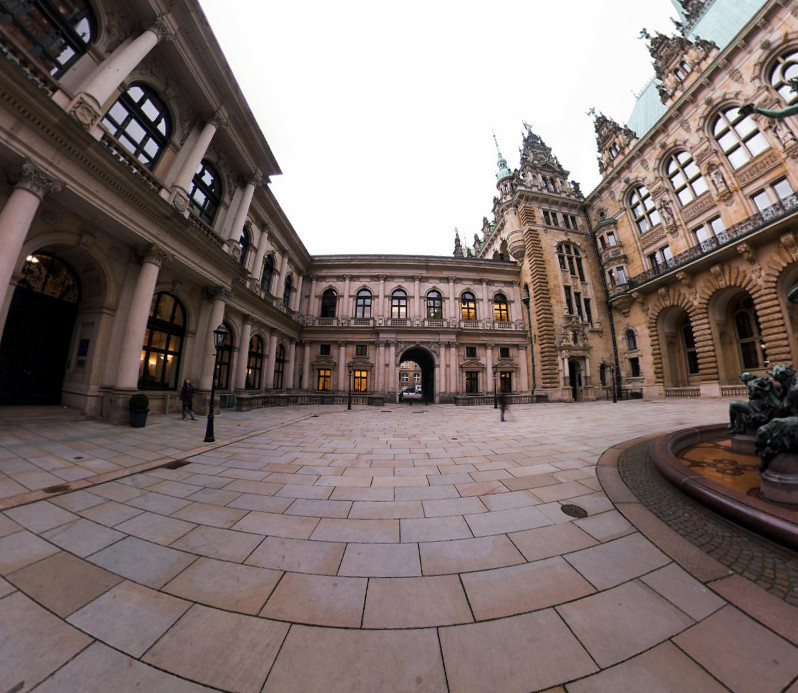}
    &
    \includegraphics[height=4.5cm]{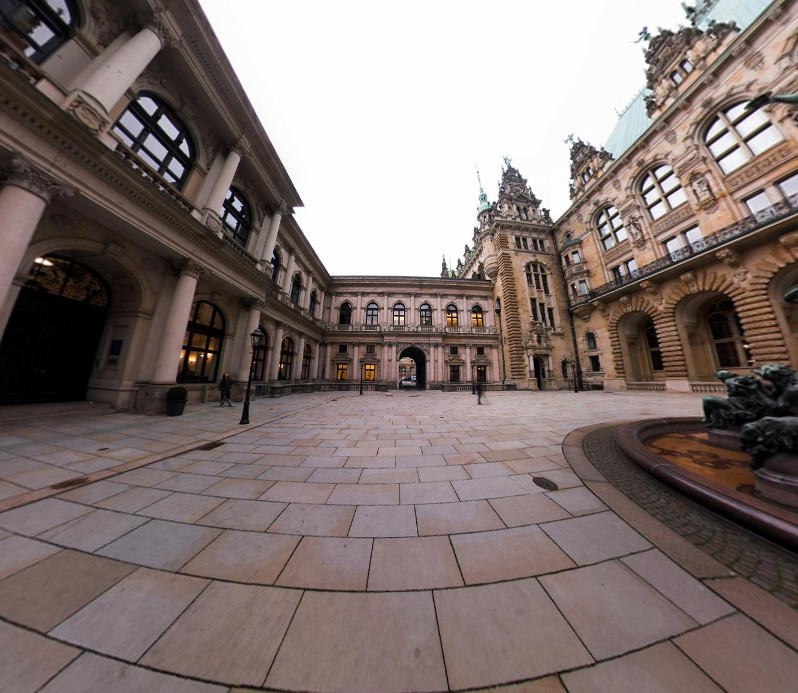}
    \\
    {\small (c) M{\"o}bius, \(\phi_\text{max}=60\degree\)}
    &
    {\small (d) M{\"o}bius, \(\phi_\text{max}=110\degree
    \)}
  \end{tabular*}
  \caption{Comparison with recent methods for $\phi = 150 \degree$. In the
  result obtained by the method in subfigure~(a)~\cite{zm05} the different
  perspective projections used for different areas of the image appear
  clear and unpleasant on the floor of the scene. The method in
  subfigure~(b)~\cite{caa09} preserves all straight lines marked by the
  user, but fails to preserve the ones on the floor (which are too many to
  be marked by the user). Note also that marking a large number of lines
  would impose too many constraints on the optimization process and could
  compromise the final result. Subfigures~(c)~and~(d) were produced with
  our method. In subfigure~(c) we use the value of \(\phi_\text{max}\) we
  consider perceptually good. Subfigure~(d) shows how the bending of lines
  on the floor was slightly corrected by augmenting the value of
  \(\phi_\text{max}\), but without obtaining a perceptually better image.}
  \label{fig:comparison_state}
\end{figure}

In Figure~\ref{fig:comparison_state} we compare our method with recent
works on the same topic with a view of the courtyard of the Hamburg town
hall. The result obtained by the technique proposed by
Zelnik-Manor~\etal~\cite{zm05} (Figure~\ref{fig:comparison_state}a) shows
discontinuities on the floor produced by using different projections for
different areas of the image.  This strategy works successfully for the
building in the image, since these discontinuities are hidden by natural
orientation discontinuities in the scene, but it fails to fit the geometry
of the floor. In Figure~\ref{fig:comparison_state}b we show a result
produced by our implementation of the technique by
Carroll~\etal~\cite{caa09}. All straight lines specified by the user (see
their work for more details) are well-preserved, but the lines on the floor
appear bent, since they are too many to be marked by the user. Although
their energy minimization formulation guarantees conformality and
smoothness of the final result, it has the problem of taking some seconds
to be performed. Our method (Figures~\ref{fig:comparison_state}c~and~d),
does not rely on heavy user interaction nor on any optimization and is not
restricted to scenes with any particular geometry.

\begin{figure}
\centering
\begin{subfigure}{.325\textwidth}
\includegraphics[width=\textwidth]{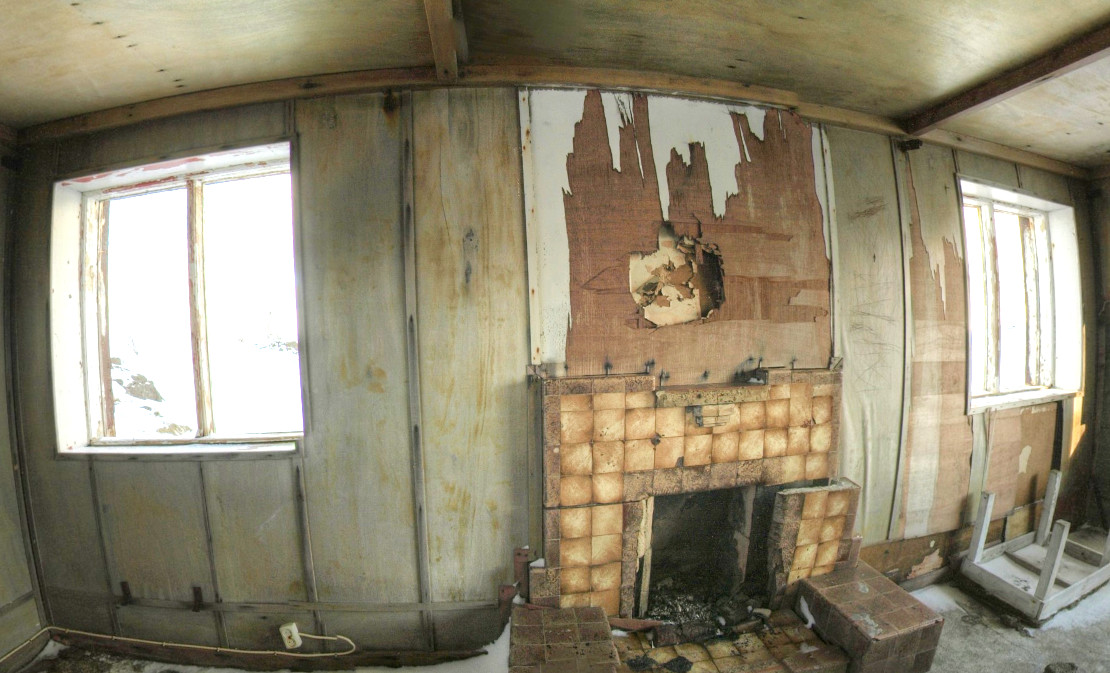}
\caption{\(\phi_\text{max}=1\degree\)}
\label{subfig:120fovmax01}
\end{subfigure}
\begin{subfigure}{.325\textwidth}
\includegraphics[width=\textwidth]{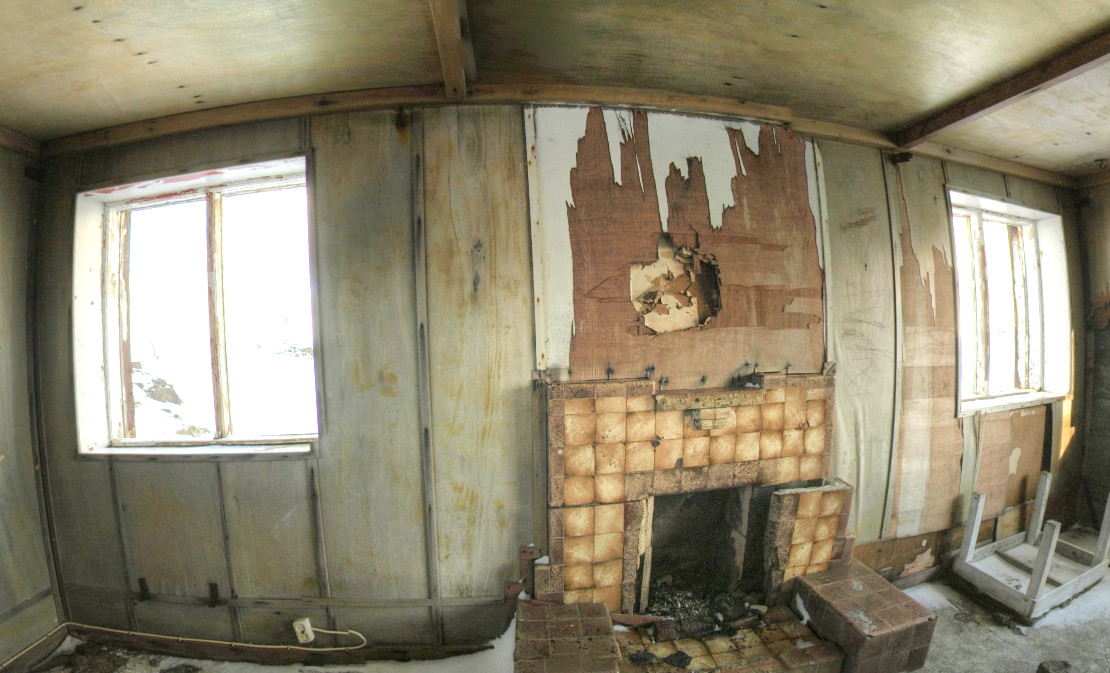}
\caption{\(\phi_\text{max}=30\degree\)}
\label{subfig:120fovmax60}
\end{subfigure}
\begin{subfigure}{.325\textwidth}
\includegraphics[width=\textwidth]{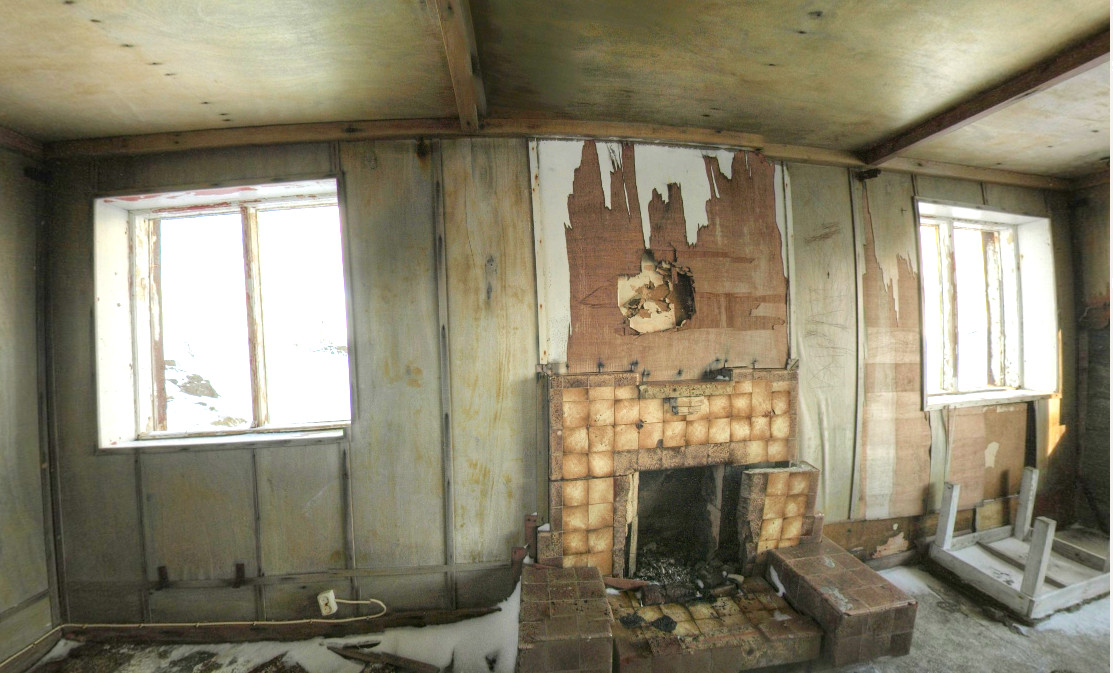}
\caption{\(\phi_\text{max}=90\degree\)}
\label{subfig:120fovmax90}
\end{subfigure}
\caption{M{\"o}bius, \(\phi=120\degree\)}
\label{fig:mobius120}
\end{figure}

\begin{figure}
\centering
\begin{subfigure}{.325\textwidth}
\includegraphics[width=\textwidth]{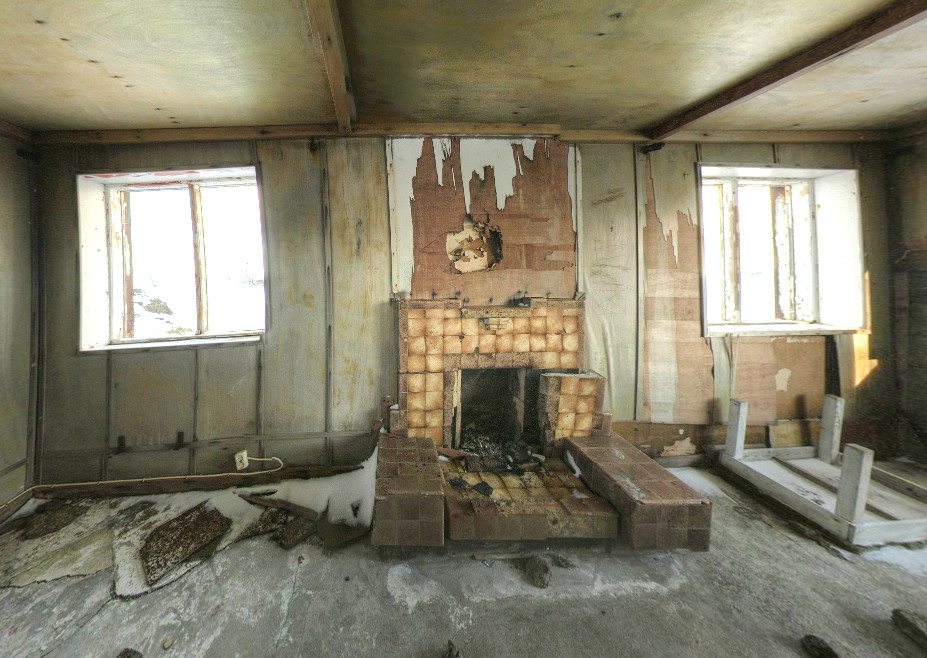}
\caption{\(d=0\)}
\label{subfig:p000}
\end{subfigure}
\begin{subfigure}{.325\textwidth}
\includegraphics[width=\textwidth]{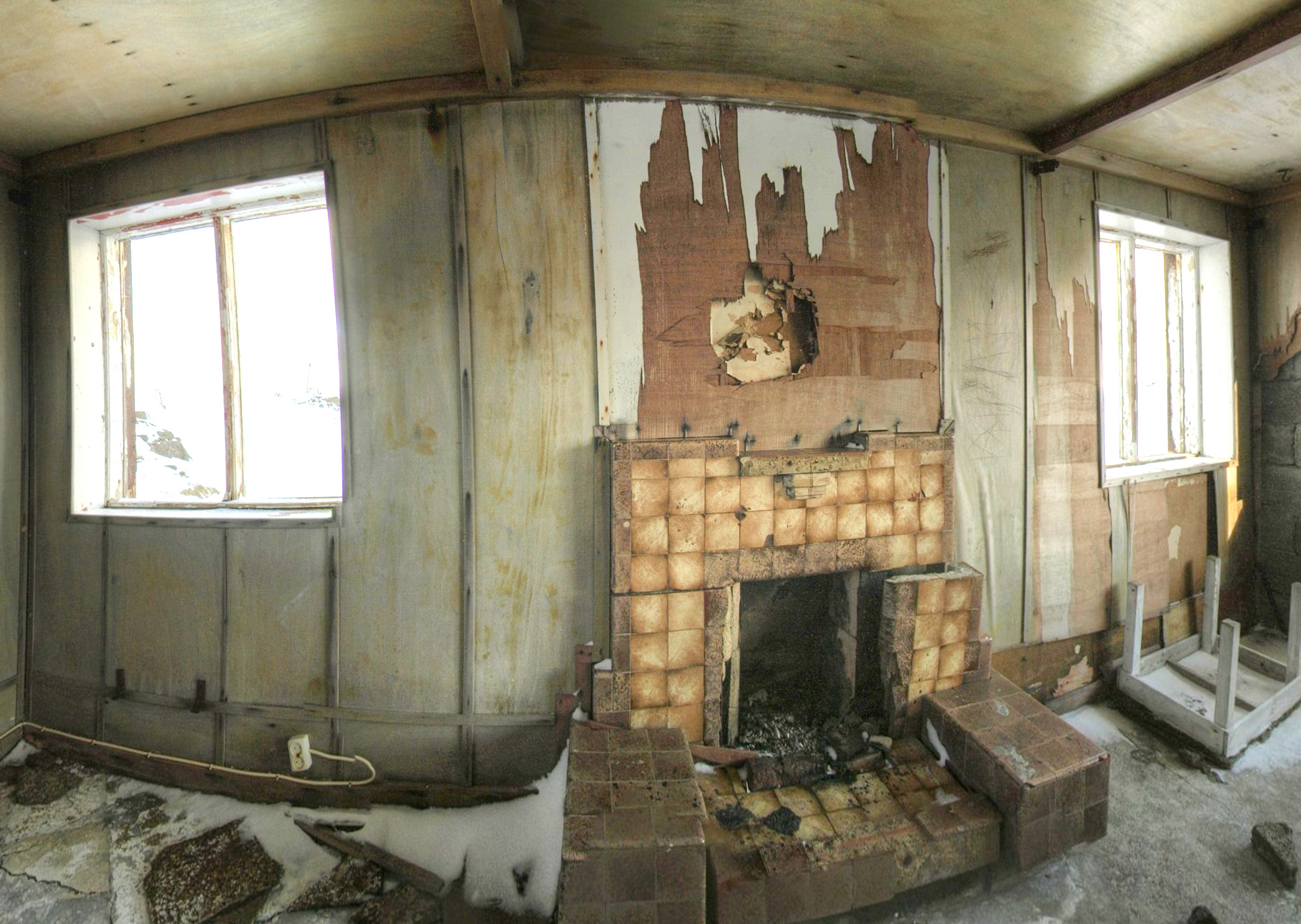}
\caption{\(d=1\)}
\label{subfig:p100}
\end{subfigure}
\begin{subfigure}{.325\textwidth}
\includegraphics[width=\textwidth]{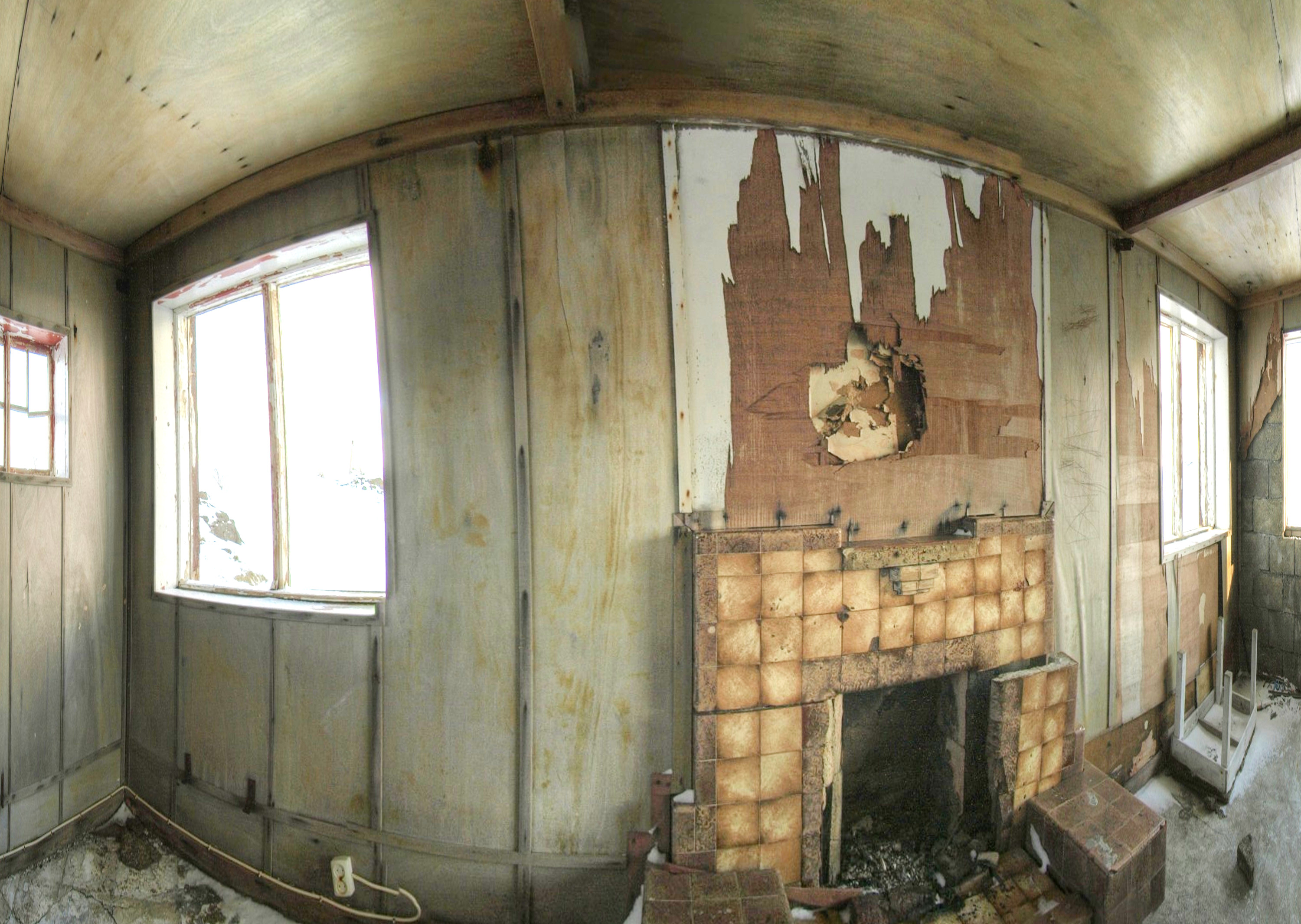}
\caption{\(d \rightarrow \infty\)}
\label{subfig:p120}
\end{subfigure}
\caption{Pannini, \(\phi=120\degree\)}
\label{fig:pannini120}
\end{figure}

Finally, we compare our approach with the Pannini projection. This
projection method works as a cylindrical projection, but moving the
projection center away from the center of the sphere. The projection center
\(c\) is moved along the ray originating at the midpoint \(o\) of the
cylinder central axis, in a direction perpendicular to the projected plane.
Assuming the cylinder has radius \(1\), the parameter \(d\) represents the
distance from \(o\) to \(c\). This parameter \(d\) controls the image
\emph{compression}. If \(d=0\) (\ie, \(o\) and \(c\) coincide), the
resulting projection is a central cylindrical projection. If \(d=1\)
(\ie, \(c\) is on the surface of the cylinder, behind \(o\)),
the projection is stereographic cylindrical (also called \emph{basic
Pannini}). When \(d \rightarrow \infty\), the resulting projection is
cylindrical orthographic. This way, when \(d\) grows, points in the borders
of the resulting image are not distorted as in the perspective projection,
producing perceptually pleasant results. Since it is a cylindrical
projection, Pannini has the property of preserving all vertical lines.

Figures~\ref{fig:mobius120}~and~\ref{fig:pannini120} show the effect of
applying our method and the Pannini projection, both with different
paramenters, to an image of a room with a FOV of
\(120\degree\)
(the latter produced using an open-source implementation of the Pannini
method, provided by the Panotools software~\cite{ptools} called through
Hugin~\cite{hugin}; please note that the Pannini projection was also
implemented by its authors as a standalone package~\cite{pvqt}).
These two figures show the effect on line
bending obtained by varying the value of the parameters of each method.
We note first one of the strengths of the Pannini projection, the
preservation of vertical lines.
For the rest, it can be considered that \(\phi_\text{max}\) and
\(d\) are somewhat analogous.
We can also observe that big values of \(d\) allow for more line bending
than small values of \(\phi_\text{max}\), but this fact does not mean that
images obtained with big values of \(d\) are perceptually better.

\begin{figure}
\centering
\begin{subfigure}{.49\textwidth}
\centering
\includegraphics[width=\textwidth]{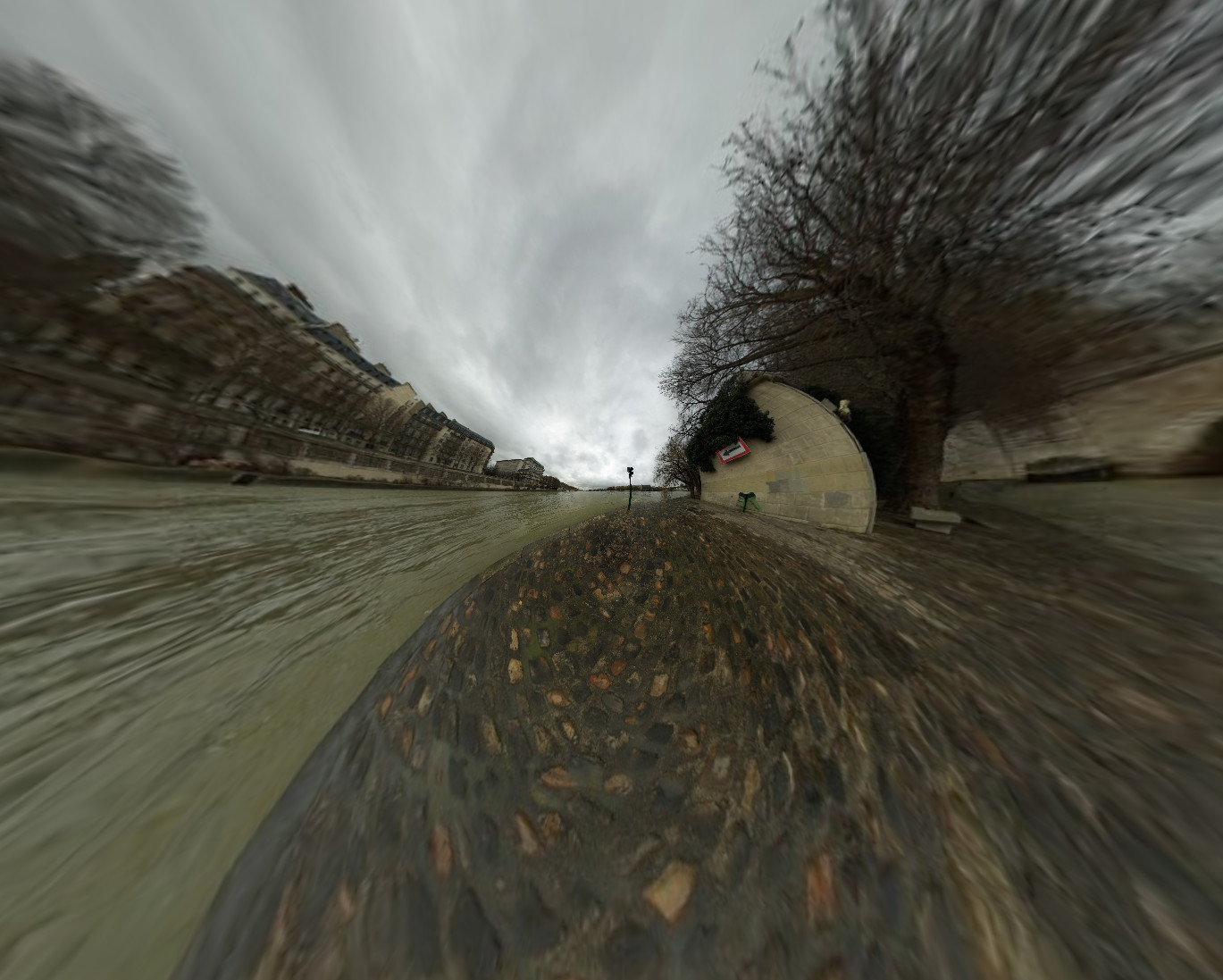}
\caption{\(\phi_{\text{max}}=40\degree \approx 0.7\text{rad}\)}
\end{subfigure}
\begin{subfigure}{.49\textwidth}
\centering
\includegraphics[width=\textwidth]{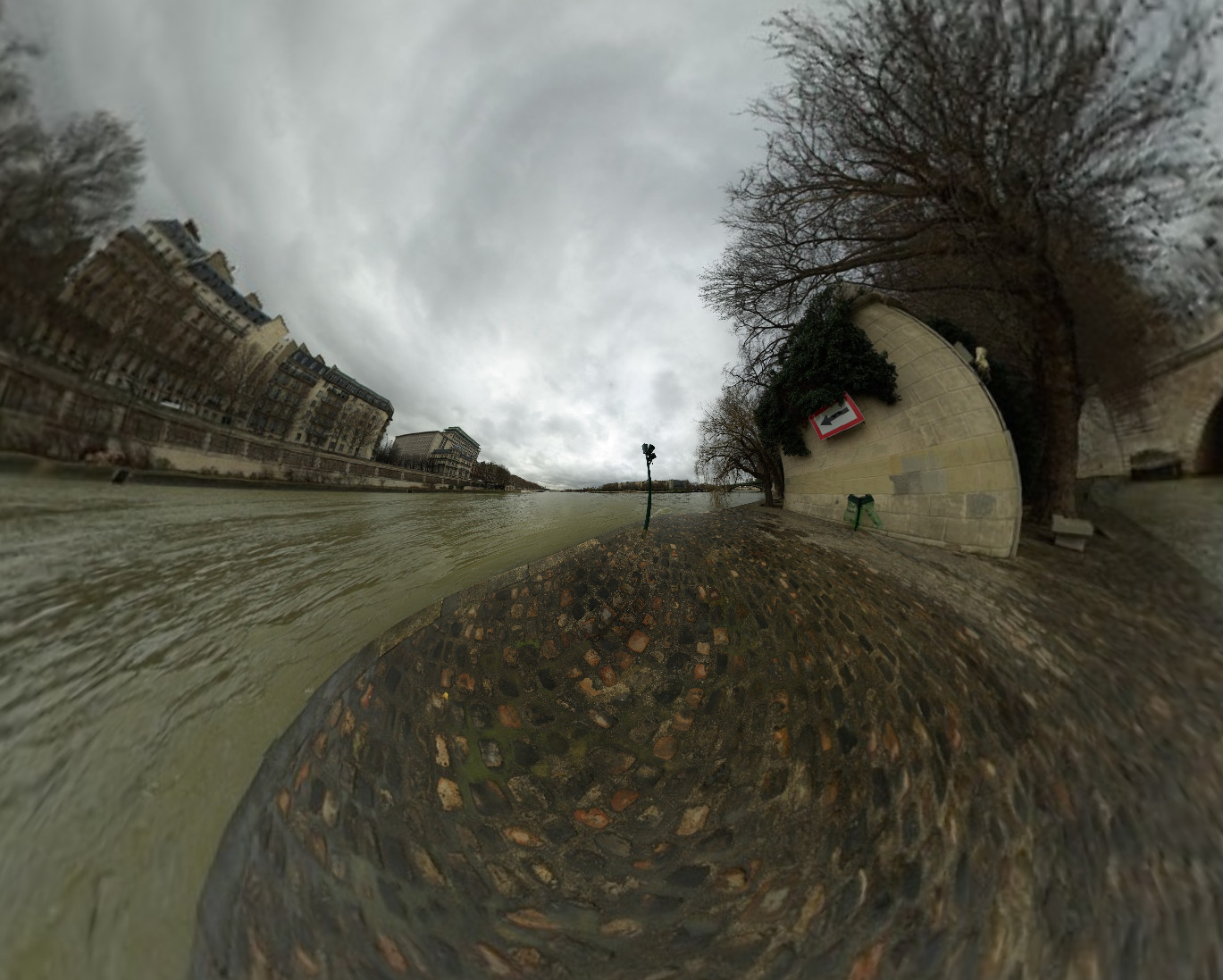}
\caption{\(\phi_\text{max}=1\degree \approx 0.02\text{rad}\)}
\end{subfigure}
\caption{Even if our method can handle very wide FOVs and displayed objects
are distinguishable, the image seems unrealistic. Here, \(\phi=320\degree
\approx5.6\text{rad}\).}
\label{fig:fov320}
\end{figure}

To conclude this Section, we recall that our method produced very good
results for all the tested cases, and was succesfully compared to
state-of-the-art projection and warping methods. Nevertheless, we should
point out that there are still some cases for whose our method cannot
produce realistic results. For instance, Figure~\ref{fig:fov320} shows what
happens with a very wide FOV, close to \(2\pi\text{rad}=360\degree\), in a
panoramic taken from the {\^I}le~Saint-Louis in Paris. A perspective
projection will simply not work in this case. Our method can produce a
clear but unrealistic image. We believe that this cannot be overhauled,
since the human vision is unable to observe such a wide FOV. The fact that
our method handles such a large FOV makes it a candidate for visualization
in different settings, such as those enumerated in the next Section.

\section{Applications}
\label{sec:appl}

The method described above does not depend on the content on the image, nor
require human interaction to specify which regions of the image should be
preserved (it is a global projection, not a local warping technique). The processing
time needed is thus much smaller, making it suitable to be applied on
panoramic videos and interactive panorama visualizers, such as Google
Street View~\cite{gsv}.

In interactive applications, apart from the resulting image, the response
speed of the interface is very important. One improvement that can be
incorporated to our technique is to have various input spheres (in
practice, various equirectangular images), with different resolutions. This
would allow to choose one input sphere, based on the current FOV,
preventing the projection to process unnecessarily large amounts of points.
The values of \(\phi\) for which the sphere must be switched must be
estimated experimentally.

Note that actually something similar to the latter approach is implemented
in Google Street View where, from one input sphere, multiple projections
with different resolutions are produced~\cite{adffllow10}; the viewing
sphere is never downloaded (this also minimizes the amount of data
transmitted, which is another constraint to keep in mind when working on a
client/server environment).

Our method can be coupled with common user interfaces of panorama
visualizers. For instance, movements with the mouse when holding a button
pressed can be used to pan the image (this is, to modify \(\alpha\) and
\(\lambda\)) and the mouse wheel can be used to increase and decrease the
FOV (that is, varying at the same time \(\phi_\alpha\) and
\(\phi_\lambda\)). These interactions can be alternatively done with keys
(for instance, panning with cursor arrows while changing the FOV with
page~up and page~down
keys).

Since the method is not computationally expensive, considering current
technological developments, it can be implemented by processing
transformations on a GPU, using a shader. For instance, the image can be a
texture to be applied on the surface of the viewing sphere. In this
setting, when panning, neither the sphere nor the texture are altered. When
zooming, the texture has to be updated in order to reflect the new
M{\"o}bius transformation applied.
We showed that applying M{\"o}bius transformations in the GPU, using GLSL,
is viable and yields high frame rates (see Table~\ref{table:performance}).

It is important to stress that, in case of working through a client/server
environment, the processing can be easily implemented in the
client (with local GPU computations as described in the previous
paragraph), in the server or even with a hybrid approach.

To conclude the present Section, let us mention projection on a dome,
another important application of our method. This technique consists in
projecting an image into the interior surface of a sphere, or a portion of
a sphere. Historically, domes where used in planetariums: they showed the
stars exactly as we see them. In the last decades, researchers realized
that domes could produce very pleasant visual experience not only when
visualizing stars; a new research direction in Computer Graphics born. A
dome can provide an immersive environment for movies, display of data or
games. While specialized hardware tends to be expensive, there are modern
low-cost approaches which make domes very appealing for non-professional
uses~\cite{b05,bf10}.
Since the projection surface on a dome is not plane, plane projection
techniques, which evolved since man started to draw maps, are useless.
Dome projection methods must consider a sphere, or a portion of it, as
input, and must project on the surface of the sphere. Points on the input
sphere must be mapped onto the output sphere.
In this setting, a M{\"o}bius transformation shall be used to directly
shrink (or, in this case, also expand) the sphere to displace points to the
visible part of the sphere, without using the last perspective projection
step, obtaining thus a plane-projection-less dome visualization.

\section{Conclusions, Discussion and Future Work}
\label{sec:future}

We introduced a technique based on M{\"o}bius transformations of the points
on the viewing sphere, aimed at improving the perceptual quality of
panoramas.
We validated the new approach by implementing the method, to demonstrate
both its quality and the speed at which it can run.

The methods by Zorin and Barr~\cite{zb95}, by Carroll~\etal~\cite{caa09}
or by Wei~\etal~\cite{wlhmt12}
consist in the minimization of energies, which can be used to quantify the
perceptual quality of the image but, of course, they need human
interaction. Milnor's method was presented more than 40 years ago, and it
was conceived to help in a slightly different problem than ours (map
projections). This fact motivates us to propose a future research
direction: the development of theoretical methods aimed at quantifying the
quality of projections without user interaction.

Throughout the paper, we use a constant value to estimate a good value for
\(\phi_\text{max}\), under the assumption that this value is good on the
perceptual side. Another interesting research direction would be to study
the content of the images, in order to determine the value of
\(\phi_\text{max}\) in function of the topology of the objects which are
present in the image. This research may also provide another approach to
measure automatically the perceptual quality of projections. Nevertheless,
our method can lose the property of being executed in real-time. A related
possible research direction would be to study the content of different
frames in a \(360\degree\) video, in order to dynamically change the
projection on each frame. This research could also include comparisons with
state-of-the-art methods for video correction~\cite{wlhmt12}.

As mentioned in Section~\ref{sec:appl}, our method can be used to visualize
\(360\degree\) images on a dome. This has very important implications that
should be discussed here. It follows that the well-studied methods that
project the viewing sphere onto a plane or a cylinder are not useful on this
setting. Thus, our method can be viewed as a member of another family of
transformations on the sphere. We also mention that our method can be (and
actually was) implemented in real-time, what permits to apply it on
\(360\degree\) videos.

A topic related to dome projection, out of the scope of the paper, which
can be considered as an application of our method, is zooming images. The
idea is simple when working on plane images or videos: zooming consists in
narrowing or widening the FOV. However, this is no longer true in
\(360\degree\) images or videos, because the FOV cannot be adjusted. A zoom
on a sphere consists in a transformation, shrinking some region but,
naturally, stretching another region. This is exactly what the M{\"o}bius
transformations presented in this paper do. Thus, the transformation of the
viewing sphere we proposed throughout the paper is a zooming technique for
\(360\degree\) images.

Finally, let us mention that M{\"o}bius transformations are a simple and
well-known mathematical tool which was not deeply explored in Computer
Graphics and Image Processing. As mentioned earlier, we believe their
applications can be extended beyond panorama visualization, namely, where
an image is represented on a viewing sphere.  This is an interesting
direction to explore in the future. To uncover applications of M{\"o}bius
transformations in visualization, it would be interesting to make a deeper
study of the foundations of this method.  Exploring notions such as the
cross-ratio~\cite{n97} can shed light to a myriad of new techniques for the
transformation of images inscribed on the viewing sphere, such as changing
the perspective of scenes with M{\"o}bius transformations.

It also remains open the question about the perceptual quality of
projections: is it possible to conceive an analytic method to quantify the
quality in a way that is consistent with human perception?

\paragraph{Acknowledgements:}
The authors thank the Flickr and Wikimedia~Commons users who made available
their equirectangular images under the Creative Commons license, used to
obtain some figures of the paper: Gadl (Figures~\ref{fig:fov240},
\ref{fig:mobius120}, \ref{fig:pannini120} and~\ref{fig:fov320}), Luca~Biada
(Figure~\ref{fig:fov172}),
DXR (Figure~\ref{fig:comparison_standard})
and HamburgerJung (Figure~\ref{fig:comparison_state}).
L.~Sacht acknowledges the doctoral scholarship from CNPq.
L.~Pe{\~n}aranda acknowledges financial support from IMPA during years
2012 to 2014.

\bibliographystyle{plain}
\bibliography{refpsv}

\newpage
\appendix
\section{M{\"o}bius Transformations}
\label{app:mobius}

We recall in this appendix some properties of M{\"o}bius transformations.
The information here should be enough to understand the paper. We refer
to the book by Needham~\cite{n97} for proofs and details.

A M{\"o}bius transformation is a mapping of the form
\[
M(z)=\frac{az+b}{cz+d}\text{,}
\]
where \(a\), \(b\), \(c\) and \(d\) are complex constants. Multiplying
these coefficients by a constant yields the same mapping, thus what matter
are the ratios of the coefficients (usually the coefficients are normalized
by scaling them to satisfy \(ad-bc=1\)). This fact shows that only three
complex numbers are sufficient to determine uniquely a mapping. It can be
shown that there exist a unique M{\"o}bius transformation which sends any
three points to any other three points.

\begin{figure}
\centering
\begin{subfigure}{.24\textwidth}
\centering
\includegraphics[width=\textwidth]{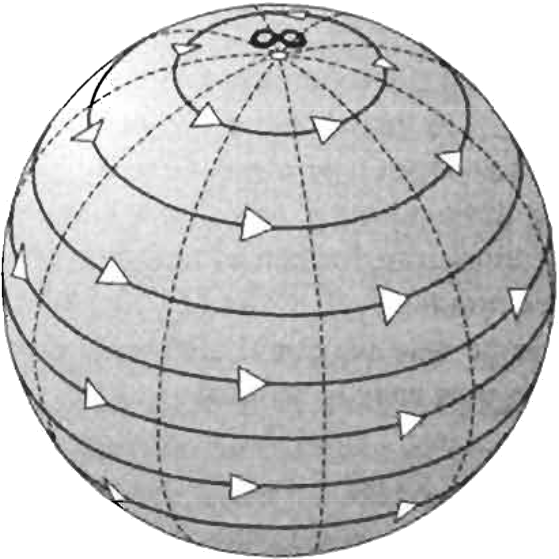}
\caption{elliptic}
\label{subfig:elliptic}
\end{subfigure}
\begin{subfigure}{.24\textwidth}
\centering
\includegraphics[width=\textwidth]{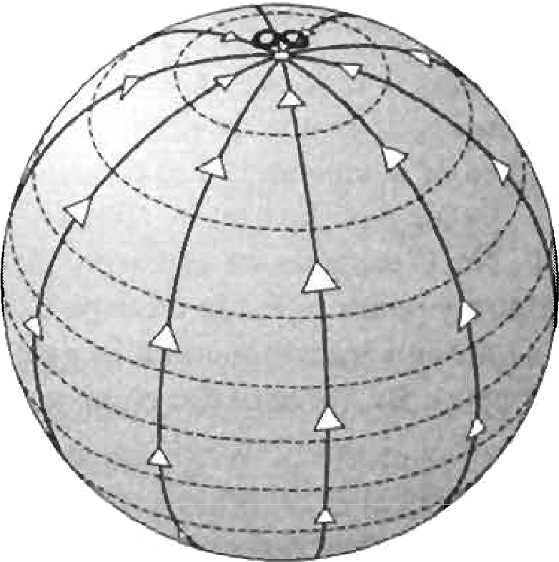}
\caption{hyperbolic}
\label{subfig:hyperbolic}
\end{subfigure}
\begin{subfigure}{.24\textwidth}
\centering
\includegraphics[width=\textwidth]{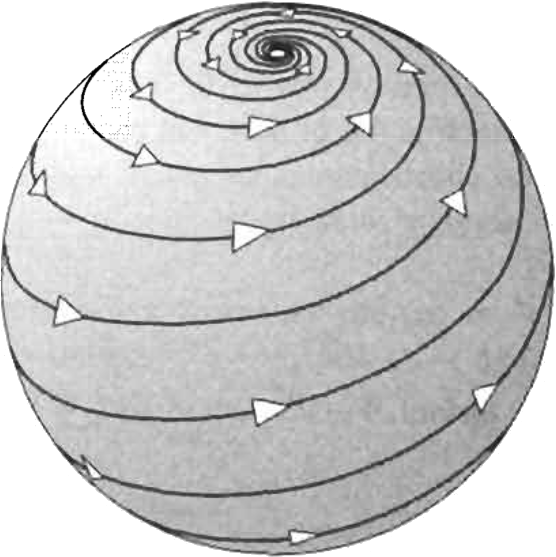}
\caption{loxodromic}
\label{subfig:loxodromic}
\end{subfigure}
\begin{subfigure}{.24\textwidth}
\centering
\includegraphics[width=\textwidth]{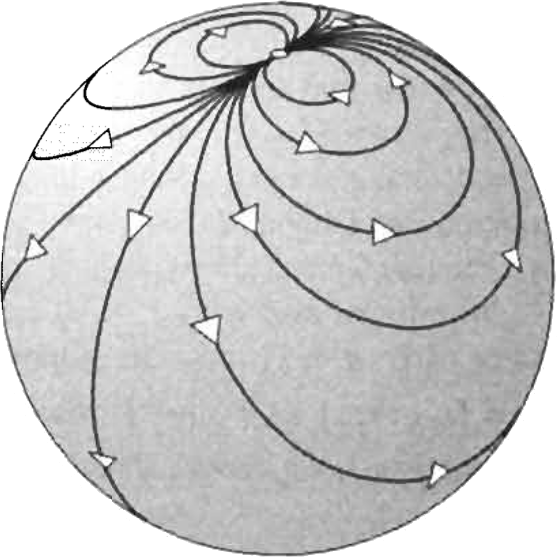}
\caption{parabolic}
\label{subfig:parabolic}
\end{subfigure}
\caption{Four kinds of M{\"o}bius transformations (drawings taken
from~\cite{n97}).}
\label{fig:kinds}
\end{figure}

M{\"o}bius transformations have fixed points, computed through the equation
\(z=M(z)\). Since this equation is quadratic, it has at most two solutions
(except for the identity mapping). When \(c\ne 0\), both fixed points lie
in the finite plane. When \(c=0\), at least one of the fixed points lies at
infinity (as it can be deduced from from Figure~\ref{fig:kinds}, infinity
is represented by the north pole \(N\) of the Riemann sphere). In this
case, the transformation is simplified, taking the form \(M(z)=Az+B\). We
can write \(A=\rho e^{i\alpha}\) in order to view \(M(z)\) as the
composition of a rotation of \(\alpha\) centered in the origin, plus an
expansion by \(\rho\) and a translation of \(B\). This interpretation lets
us view geometrically the M{\"o}bius transformation, as depicted in
Figure~\ref{fig:kinds}.

When \(\alpha>0\), \(\rho=1\) and \(B=0\), \(M(z)\) is a rotation of the
complex plane (\ie, a rotation of the sphere), as shown if
Figure~\ref{subfig:elliptic}. The fixed points of this transformation are
the two poles of the sphere, which correspond, in the complex plane, to the
origin and infinity. This is called an \emph{elliptic} M{\"o}bius
transformation.

Figure~\ref{subfig:hyperbolic} illustrates the transformation when
\(\alpha=0\), \(\rho>1\) and \(B=0\). \(M(z)\) is, in this case, an
expansion centered in the origin. The two fixed points are the same as in
the previous case. If \(\alpha=0\), \(\rho<1\) and \(B=0\), it is an
origin-centered contraction. These M{\"o}bius transformations are called
\emph{hyperbolic}.

When \(\alpha\ne 0\), \(\rho\ne 1\) and \(B=0\), the resulting
transformation is a combination of the two former cases. This is called a
\emph{loxodromic} M{\"o}bius transform and it is illustrated in
Figure~\ref{subfig:loxodromic}. The two fixed points of this transform are
the same two as the above cases.

The last case study is the translation, occurring when \(A=0\) and \(B\ne
0\). The only fixed point of this transformation is infinity, represented
on the sphere by the north pole. This transformation, the \emph{parabolic}
M{\"o}bius transformation, is depicted in Figure~\ref{subfig:parabolic}.

\end{document}